\documentclass[journal]{IEEEtran}

\usepackage[noadjust]{cite}
\usepackage[utf8]{inputenc}
\usepackage[export]{adjustbox}
\usepackage{subfig}
\usepackage{xcolor,colortbl}
\usepackage{pstricks-add} 
\usepackage{tikz,pgfplots}
\usetikzlibrary{patterns}
\usetikzlibrary{positioning}
\usepackage{balance}
\usepackage{booktabs} 
\usepackage{amsmath,amssymb,amsfonts}
\usepackage{algorithm}
\usepackage{algpseudocode}
\usepackage{braket}
\usepackage{graphicx}
\usepackage{textcomp}
\usepackage{fixltx2e}
\usepackage[hyphenbreaks]{breakurl}
\usepackage{svg}
\usepackage{amsmath, amsthm}
\usepackage{pgfplots}
\usepackage{pgfplotstable}
\usepackage{subfig}
\usetikzlibrary{pgfplots.groupplots}
\usepackage{soul}
\usepackage{caption}
\usepackage{listings}
\usepackage{mathtools}
\usepackage{graphics}
\usepackage{enumitem}
\usepackage[printwatermark]{xwatermark}
\usepackage{makecell}
\usepackage{svg}
\usepackage{pst-plot}    
\usepackage{xcolor}
\usepackage{algorithm}
\usepackage{algpseudocode}%
\usetikzlibrary{patterns}
\usetikzlibrary{positioning}
\usepackage{gensymb}
\usepackage{float}
\usepackage{bm}
\usepackage{array}
\newcolumntype{?}{!{\vrule width 1pt}}
\usepackage[draft]{hyperref}
\newcommand{\myindent}[1]{
	\newline\makebox[#1cm]{}
}

\usepackage[flushleft]{threeparttable}

\usepackage[all]{background}
\usepackage{stackengine}
\setstackEOL{\\}
\setstackgap{L}{\normalbaselineskip}
\SetBgContents{\color{blue}{\tiny \Longstack{PREPRINT - accepted at IEEE Transactions on Computer-Aided Design of Integrated Circuits and Systems (TCAD), 2021.}}}
\SetBgPosition{3.5cm,1cm}
\SetBgOpacity{1.0}
\SetBgAngle{0}
\SetBgScale{1.8}

\algnewcommand{\LineComment}[1]{\State \textcolor{blue!70}{/* #1 */}}


\definecolor{ForestGreen}{RGB}{34,139,34} 
\definecolor{Gold}{RGB}{218,165,32} 
\definecolor{MediumBlue}{RGB}{25,25,205} 
\definecolor{Plum}{RGB}{186,119,183} 
\definecolor{bb1}{RGB}{178,24,43}
\definecolor{bb2}{RGB}{244,165,130}
\definecolor{bb3}{RGB}{33,102,172}

\definecolor{c1}{RGB}{171, 171, 171}
\definecolor{c2}{RGB}{217,217,217}
\definecolor{c3}{RGB}{60,179,113}
\usetikzlibrary{patterns}

\definecolor{Cgray}{RGB}{101,104,124}
\definecolor{Cblue}{RGB}{63,136,197}
\definecolor{Cred}{RGB}{244,44,67}
\definecolor{Cgreen}{RGB}{46,160,92}

\definecolor{b0}{RGB}{255,153,153}
\definecolor{b1}{RGB}{235,56,56}
\definecolor{b2}{RGB}{191,0,0}
\definecolor{b2}{RGB}{200,0,0}
\definecolor{b3}{RGB}{100,151,177}
\definecolor{b4}{RGB}{0,91,150}
\definecolor{b5}{RGB}{1,31,75}
\definecolor{gr}{RGB}{102, 102, 102}

\definecolor{orange}{RGB}{250, 154, 80}

\definecolor{rwth1blue}{RGB}{0, 84, 159}
\definecolor{rwth2orange}{RGB}{246, 168, 0}
\definecolor{rwth2gray}{RGB}{100, 101, 103}
\definecolor{rwth2red}{RGB}{204,7,30}

\DeclarePairedDelimiter\abs{\lvert}{\rvert}
\ifCLASSINFOpdf
\else
\fi

\begin{document}
\bstctlcite{IEEEexample:BSTcontrol}
%
\title{Deceptive Logic Locking for Hardware Integrity Protection against Machine Learning Attacks}
%
%
%
\author{Dominik~Sisejkovic,
	Farhad~Merchant,
	Lennart M. Reimann,
	and~Rainer~Leupers
	\thanks{Dominik~Sisejkovic,	Farhad~Merchant, Lennart M. Reimann, and~Rainer~Leupers are with RWTH Aachen University, 52062 Aachen, Germany, e-mail:  \{sisejkovic, merchantf, reimannl, leupers\}@ice.rwth-aachen.de.}
	\thanks{}}

%
%

{}
%



\maketitle

\begin{abstract}
Logic locking has emerged as a prominent key-driven technique to protect the integrity of integrated circuits. 
However, novel machine-learning-based attacks have recently been introduced to challenge the security foundations of locking schemes. These attacks are able to recover a significant percentage of the key without having access to an activated circuit. This paper address this issue through two focal points. First, we present a theoretical model to test locking schemes for key-related structural leakage that can be exploited by machine learning. 
Second, based on the theoretical model, we introduce D-MUX: a deceptive multiplexer-based logic-locking scheme that is resilient against structure-exploiting machine learning attacks.
Through the design of D-MUX, we uncover a major fallacy in existing multiplexer-based locking schemes in the form of a structural-analysis attack. Finally, an extensive cost evaluation of D-MUX is presented.
To the best of our knowledge, D-MUX is the first machine-learning-resilient locking scheme capable of protecting against all known learning-based attacks. Hereby, the presented work offers a starting point for the design and evaluation of future-generation logic locking in the era of machine learning.
\end{abstract}


%
\IEEEpeerreviewmaketitle

\section{Introduction}\label{introduction}

Logic Locking (LL) is a premier technique to safeguard the integrity of hardware designs throughout the Integrated Circuit (IC) supply chain~\cite{evo2017,yasin2020trustworthy}. This obfuscation technique performs functional and structural design alterations through the insertion of key-dependent logic. 
Nevertheless, the security of LL has been challenged in the past decade through different key-recovery attacks. While all attacks assume the existence of the design in locked netlist format, the differentiating factor is the availability of an \textit{oracle}, i.e., an activated IC instance. The oracle is used to acquire golden Input/Output (I/O) patterns. Thus, all attacks can be classified into Oracle-Guided (OG)~\cite{decadeOfLocking} and Oracle-Less (OL) attacks~\cite{sweepAttack2019,TGA2020,desynthesisAttack2017,redundancyIdentification2019,sisejkovic2020challenging}. The OG model is manifested in high-volume commercial fabrication, where it is assumed that an attacker is able to purchase an unlocked instance of the design~\cite{pilato2020assure}. In comparison, the OL scenario is relevant in a low-volume setup where an attacker is not able to get a copy of the activated IC. This is, for example, the case in the development of security-critical systems with unique and highly confidential hardware requirements~\cite{force2005high, sisejkovic2020scopes}. 

Moreover, recent findings strongly suggest that the OL model plays a more important role in realistic attack scenarios. These include the following.
$(i)$ The secret key can be exposed by hardware Trojans that leak the key once the IC is activated regardless of the LL scheme~\cite{TAAL2019}. $(ii)$ Various OG attacks rely on having access to the internal states of a design; otherwise, the attacks become impractical. However, this assumption is unrealistic, since legitimate IC vendors never leave a scan-chain open and typically use some form of authentication~\cite{secureScanChain}. $(iii)$ Recent works have demonstrated the extraction of keys from activated ICs even in the presence of a tamper- and read-proof memory through probing or fault-injection attacks~\cite{KeyReading2019, Engels2019TheEO,jain2020atpgguided}.
These observations exclude the necessity for some of the most efficient OG attacks as the key is retrievable from the \textit{activated} IC regardless of the LL scheme. 

Furthermore, it is important to understand the exact security implications of LL. First, it can render a fabricated IC inoperable, as the correct operation is only ensured if the activation key is provided. However, as discussed in~\cite{OnTheImpossibility2019}, achieving inoperability can be done without LL. Hence, the primary objective of LL is to ensure concealing the design's functionality. This security property is known as \emph{functional secrecy}~\cite{OnTheImpossibility2019}. In this context, even though we measure the success of key-recovery attacks in terms of key accuracy, acquiring the key itself is not the main threat model. Since the design concept of locking ensures that the key impacts the functional and structural alteration of the design, the key-accuracy metric acts as a \emph{proxy} to measuring functional secrecy. Hereby, security can be defined in the form of Exact Functional Secrecy (EFS), Approximate Functional Secrecy (AFS), and Best-Possible Approximate Functional Secrecy (BPAFS); for both the OG and OL model~\cite{OnTheImpossibility2019}. EFS embodies the security against a \emph{perfect} reconstruction of the design under attack. In that regard, AFS offers a stricter criterion; it requires \emph{approximation} resiliency, i.e., security against reconstructing a design that disagrees with the original up to a selected $\epsilon$ fraction of its input space. BPAFS represents a relaxed form of AFS, where the attacker has a certain \emph{a priori} knowledge about the circuit's size and depth. Evidently, AFS-OG implies AFS-OL and EFS-OG implies EFS-OL, respectively. Furthermore, AFS implies EFS; and BPAFS-OL is equivalent to AFS-OL.

Note that functional secrecy can be provably achieved. For example, an adaptation of the SFLL scheme is provably secure under EFS-OG/OL. However, BPAFS-OG/OL has only been achieved through universal circuits~\cite{OnTheImpossibility2019, LLTrilemma2019,sfll2017}, which suffer from impractical overheads. Therefore, achieving BPAFS-OL with traditional, low-cost LL is still an open problem. Hence, this goal remains in the main focus of this work.

Moreover, with the proliferation of Machine Learning (ML) across various domains, ML is slowly being introduced into hardware security as well. A few recent works have evaluated ML models for attacking LL~\cite{GenUnlock2019,azar2020nngsat,particleSwarmBasedLL2020,SAIL2019,sisejkovic2020challenging,BOCANet2019}. Nevertheless, so far, there has been a gap in designing LL that is resilient against ML-based attacks as well as in the theoretical means of uncovering the source of ML-exploitable leakage.



\textbf{Contributions:} To fill the theoretical and practical gap on ML resilience in an OL setting, in this work, we design a learning-resilient locking scheme---from theory to practice. Hereby, we use the term \textit{learning resilience} to evaluate schemes against learning-based attacks that capture \textit{locking-induced structural residue} that can lead to \textit{key-related information leakage}.
The contributions are as follows:
\begin{itemize}
	\item The introduction of the first theoretical concept to test learning resilience in logic locking. 
	\item The Deceptive Multiplexer (D-MUX) logic locking; a novel scheme based on MUX insertion that is resilient against existing ML-based attacks. 
	\item We introduce a novel oracle-less Structural Analysis Attack on Multiplexer-based locking (SAAM) based on a major pitfall in existing MUX-based logic locking.
	\item We empirically evaluate the resilience of D-MUX against the oracle-less attacks SAAM, SWEEP, and SnapShot.
	%
	\item We evaluate the cost of D-MUX in terms of area, power, and delay through a theoretical and empirical assessment.
	\item Finally, based on the lessons learned, we analyze the structural aspects of related work. Hereby, we identify novel structural leakage points in existing schemes.
\end{itemize}

To the best of our knowledge, this work is the first to shed light on the theoretical concepts of learning resilience in logic locking as well as propose the first empirically evaluated LL scheme that is resilient against learning-based attacks. 

This work is organized as follows. Section~\ref{preliminaries} introduces the background on LL. Learning resilience is discussed in Section~\ref{learning:resilience:concept}. The structural-analysis attack is described in Section~\ref{saam:attack}. Section~\ref{dmux:description} introduces D-MUX. The resilience and cost evaluation of D-MUX is presented in Section~\ref{resilience:eval} and Section~\ref{costeval}, respectively. The related work is summarized in Section~\ref{related-work}. Limitations and opportunities are discussed in Section~\ref{limitations}. Finally, Section~\ref{conclusion} concludes the paper.

\section{Preliminaries}\label{preliminaries}
Logic locking can be categorized into two groups: combinational and sequential~\cite{LogicObfuscationOverview2019}. Combinational LL focuses on manipulations in the combinational path of circuits, whereas sequential LL obfuscates the state transition graph of a design. In the following, we only focus on combinational LL.

\subsection{Logic Locking}\label{LLinSupplyChain}

\begin{figure}[t]
	\centering
	\subfloat[Original circuit]{
		\includegraphics[width=0.36\columnwidth]{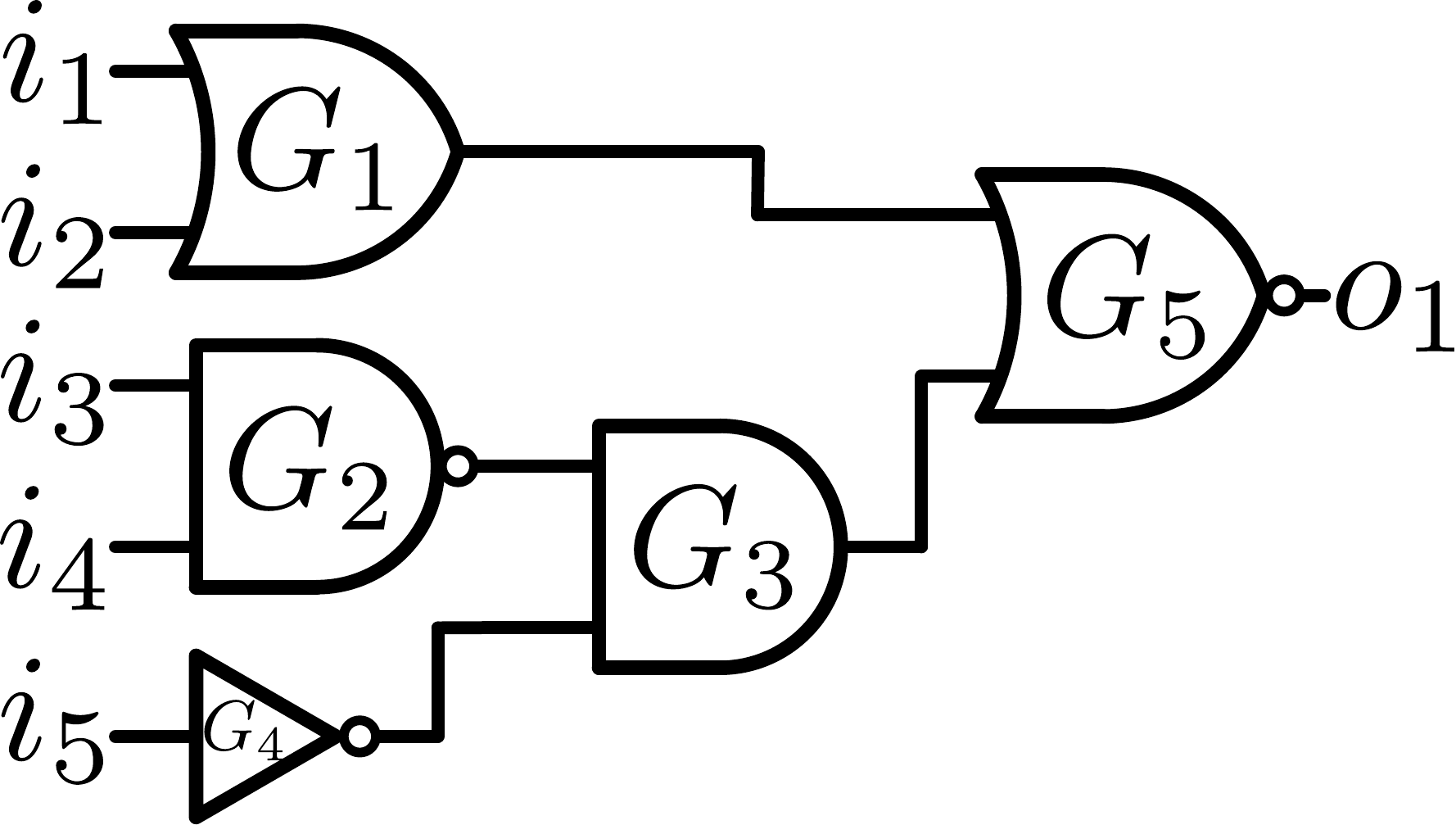}
	}
	\subfloat[Locked circuit]{
		\includegraphics[width=0.4\columnwidth]{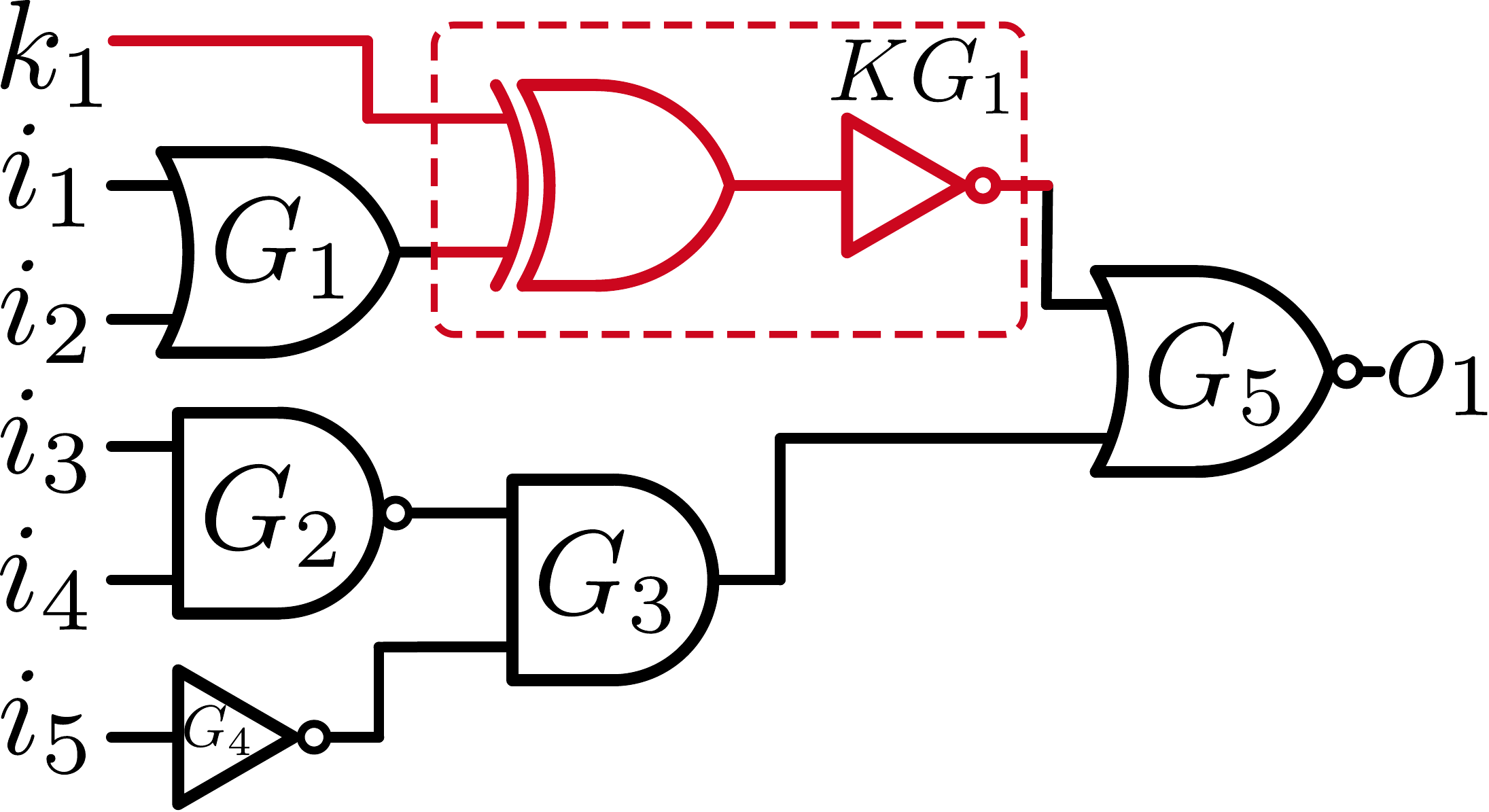}
	}
	\caption{Example: logic locking using EPIC.}
	\label{fig:enc-example}
\end{figure}

To showcase the working principles of LL, let us consider EPIC~\cite{epic2010}---one of the first LL schemes. EPIC is based on the insertion of key-controlled XOR and XNOR (XOR + INV) gates (known as \emph{key gates}) at randomly selected locations in the gate-level design. An EPIC-locked circuit is shown in Fig.~\ref{fig:enc-example}. In this example, the original circuit (Fig.~\ref{fig:enc-example}~(a)) is locked through the insertion of an XOR+INV gate bound to the key input $k_{1}$ (Fig.~\ref{fig:enc-example}~(b)). Only if $k_{1}=1$, the key gates act as a \textit{buffer}, thereby restoring the original functionality of the circuit. Otherwise, for a $k_{1}=0$, the key gates disrupt the intended functionality by \textit{inverting} the output of gate $G_{1}$, leading to faulty output values. In terms of EPIC, it is assumed that a simple removal of the key gates is prevented since an adversary must guess if the inverter is part of the key gate or the original functionality. In the past years, a wide range of combinational LL schemes has been introduced based on XOR/XNOR gates~\cite{epic2010, sll2}, multiplexers~\cite{plaza2015solving,logicConeAnalysis2015, fault2015, cyclicObf2017}, AND/OR gates~\cite{illegalOverproduction2014}, and others~\cite{sarlock2016,strongAntiSat2020, caslock2019, sfll2017}.

%


\textbf{Logic Locking in the IC Supply Chain:}~
The IP owner is the trusted entity that is introducing a legitimate product on the semiconductor market. The design is either done in-house or parts of the design services, e.g., placement and routing, are subcontracted to an external design house. In both cases, the final layout is sent to the foundry. Both the external design house and the foundry are considered untrusted entities.

The IP owner can utilize LL to conceal the design after the first logic synthesis round. Thus, LL is applied to the synthesized \textit{gate-level netlist}. After LL, the netlist is typically \textit{resynthesized} to integrate the induced changes. Hence, we can differentiate the \textit{pre-resynthesis} and \textit{post-resynthesis} netlist. 
After fabrication, the IC is returned to the IP owner for activation. The key is configured through a non-volatile memory. Since the key is only known to the IP owner, LL protects the design while in untrusted hands. Henceforth, the term \textit{netlist} refers to a gate-level netlist.



\textbf{Attack Model:} The attack model includes the following assumptions. ($i$) The adversary has only access to the locked netlist (OL model). ($ii$) The adversary is aware of the algorithmic details of the LL scheme.
($iii$) The location of the key inputs in the netlist is known. 
In the rest of this work, we refer to the netlist under attack as the \textit{target} netlist.


\subsection{Oracle-less Attacks}\label{attacksoverview}
The following provides an overview of existing OL attacks. 
Note that even though these attacks can achieve high accuracy, they do not defeat locking policies that are provably secure in the OG model~\cite{OnTheImpossibility2019}.

The desynthesis attack~\cite{desynthesisAttack2017} tries to extract the secret key through a Hill Climbing search of a series of new synthesis rounds using a randomly selected key. Hereby, the search is guided by the similarity between the locked and resynthesized netlist. 
However, this attack is not scalable to large keys and assumes a robust understanding of the synthesis tool.

The redundancy attack~\cite{redundancyIdentification2019} attempts to recover the correct key by pruning out incorrect key values when these introduce a significant level of functional redundancy in the netlist. 
Similarly, the Topology-Guided Attack (TGA) exploits the fact that basic functions in a logic cone are often repeated multiple times in a netlist~\cite{TGA2020}. Thereby, TGA can recover the correct key of a particular key gate by looking at equivalent functions that are constructed using all possible hypothesis key values.
Even though both are OL, these attacks are designed and evaluated to exploit XOR/XNOR-based LL schemes. Moreover, resilient LL against both attacks has been proposed in~\cite{redundancyShield2019, TGA2020}. 

SAIL~\cite{SAIL2019} deploys ML algorithms to retrieve the key-driven local structures of the pre-resynthesis netlist based on the post-resynthesis target. Once the reconstruction is done, SAIL extracts the correct key solely based on the properties of XOR/XNOR-based LL. Thus, it is not relevant for evaluation in this work. Moreover, recently, the UNSAIL locking technique has been proposed to thwart the SAIL attack~\cite{unsail2020}.

In the context of MUX-based locking and ML-based attacks, two attacks play a major role: \textbf{SWEEP} and \textbf{SnapShot}. Both are described in more detail in the following.

\subsubsection{SWEEP}\label{sweepdef}
This constant propagation attack exploits the differentiating circuit characteristics induced by hard-coding a single key-bit value during resynthesis~\cite{sweepAttack2019}. An overview of the subsequent steps of the attack is presented in the following. 

\textbf{Training Set Generation:}~The training set is assembled by either using a set of available locked benchmarks with known key values or by relocking the target netlist with new keys.

\textbf{Feature Extraction:}~The training data is resynthesized by hard-coding \textit{each key input} to the correct and incorrect value (logic 0 or 1). This procedure results in two synthesis reports for each key value. The relevant features are extracted from the reports and consolidated in the form of a feature matrix.

\textbf{Feature Weighting:}~In this stage, the attack generates a set of optimum weights that evaluate the correlation between the entries in the feature matrix and the known correct keys.

\textbf{Deployment:}~Finally, the design features are extracted from the target by repeating the previous steps for an unknown key. To extract the key value, SWEEP utilizes the generated weights from the previous step and the target features to deduce a value for each key input. The value can be 0, 1, or X. If X is given, the attack was not able to make a safe decision about the correct value. Moreover, SWEEP allows an adjustable margin $m$ to be selected by the attacker. $m$ controls the freedom of SWEEP to make "wild" guesses for values that are similar (close to each other). By default, $m=0$.

\textbf{Evaluation Metrics:}~SWEEP relies on two metrics: \emph{accuracy} and \emph{precision}. Accuracy is defined as the percentage of correctly extracted key-bit values out of the entire key, \textit{regardless of potential X values}, i.e., $(N_{correct}/N_{total})\cdot{100\%}$.
Precision is defined as the percentage of correct keys, where every potential X value is regarded as a correct guess, i.e., $((N_{correct} + N_{X})/N_{total})\cdot{100\%}$.


Note that SWEEP is relevant for evaluation in this work as it is an OL attack and well suited for MUX-based locking.

%

\subsubsection{SnapShot}\label{snapshotintro}
The SnapShot attack utilizes an ML model to predict key values based on structural features extracted from the target netlist~\cite{sisejkovic2020challenging}. Compared to the mentioned attacks, SnapShot has the ability to make a \textit{direct} key guess by "just" looking at the post-resynthesis netlist, without having to extract synthesis features or reconstruct the pre-resynthesis netlist. The attack flow is described in the following.

\textbf{Training Set Generation:}~Similar to SWEEP, the training set can be generated in two ways, resulting in the Generalized Set Scenario (GSS) and the Self-Referencing Scenario (SRS). In GSS, the ML model learns to predict key values based on a generalized set of locked circuits (not including the target). In SRS, a novel set is generated by copying the target multiple times and relocking each copy with an additional key.

\textbf{Extraction:}~In this step, SnapShot extracts the \emph{key-affected} netlist subgraphs that serve as training samples for the ML model. For each key input, the extraction procedure follows the key wire until a gate is encountered. This gate is considered as the central key gate. Starting from this gate, the netlist is traversed in a Breadth-First Search (BFS) fashion towards the primary I/Os. While traversing, each gate is mapped to an integer value based on an encoding table. The extracted representation is known as Locality Vector (LV). 

\textbf{ML Model Design:}~The model is trained based on the labeled LVs to predict key values for unseen, target LVs. In principle, any ML model can be used in SnapShot. However, due to the BFS-nature of the extraction, the LVs embody the structural (image) representation of the subgraphs; thus making them suitable for processing with Convolutional Neural Networks (CNNs). However, since this prediction problem is of a novel nature, the original work deploys neuroevolution to automatically design suitable CNN architectures. In addition, a shallow artificial neural network is evaluated for comparison.

\textbf{Deployment:}~Once the model is trained, SnapShot is deployed to predict the key of the locked netlist. For this task, the \textit{unlabeled} locality vectors of the target are extracted using the same procedure as for the training set. Finally, the target localities are presented to the ML model for key prediction. 

\textbf{Evaluation Metric:}~SnapShot uses the Key Prediction Accuracy (KPA) metric for attack evaluation. \textit{The KPA is equivalent to the accuracy definition used in SWEEP except that all bits are always guessed (no X values are allowed).}

Note that other attacks utilize ML-models as well; however, only in the form of OG attacks~\cite{GenUnlock2019,azar2020nngsat,particleSwarmBasedLL2020}. 

\section{The Concept of Learning Resilience}\label{learning:resilience:concept}
To evaluate LL for learning resilience, we analyze what structural changes are induced by a scheme, thereby considering two netlist variants. The first variant includes netlists that only consist of a \textit{single gate type}. The second variant covers netlists that consist of a randomly selected and well distributed amount of \textit{all gate types}. These variants represent two ends of the spectrum of possible netlist structures: regular and irregular. \textit{Note that the regularity describes the repetition of equivalent logic structures throughout the netlist}. The rationale of looking at these variants is that a structural key-related leakage is likely to repeat for similar structures in a design.
In theory, any netlist can be placed between the two variants. For example, designs that exhibit very regular and repeating structures are closer to the first variant. Examples include sbox, adder, multiplier, and decoder (tree of multiplexers) implementations. On the other hand, a very irregular structure (e.g., specific control logic) is closer to the second variant. 
The success of various attacks that exploit this regularity suggests that hardware designs are mostly regular and closer to the first case. Using this spectrum, we can devise two theoretical tests for learning resilience. The first variant is represented by the AND Netlist Test (ANT) and the second variant is represented by the Random Netlist Test (RNT). Hereby, the concept is that ANT and RNT evaluate for potential structure-related leakage at the \emph{two ends of the spectrum} (regular and irregular).



\subsection{Learning-Resilience Test} 
\begin{figure}[t]
	\centering
	\includegraphics[width=\columnwidth]{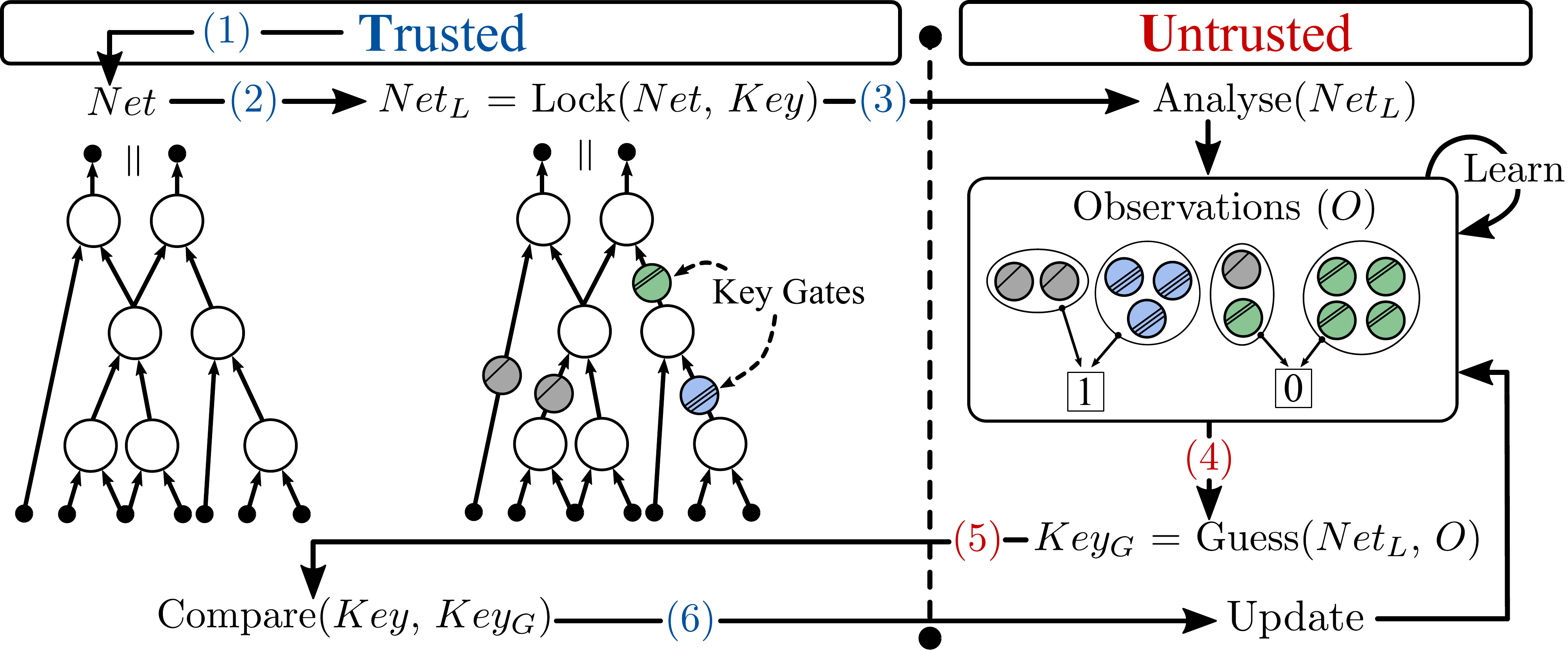}
	\caption{Learning-resilience test.}
	\label{fig:ant}
\end{figure}

The core idea of a learning-resilience test is to offer a means to uncover potential design flows of LL that lead to exploitable structural information leakage. Hereby, the test is designed around the two spectrum ends.
Both ANT and RNT follow the \textit{same test procedure}. The only difference is in the type of netlists used in the process.
The test is manifested in the form of a learning game between two parties: Trusted (\textbf{T}) and Untrusted (\textbf{U}). The goal of \textbf{U} is to learn how to predict the key based on the locked netlists provided by \textbf{T}. \textit{If a locking scheme fails at least one test, i.e., \textbf{U} \textbf{is able} to make an educated guess about the key (guessing accuracy is higher than 50\%), the locking scheme is \textbf{conclusively} \textbf{vulnerable}, otherwise it \textbf{might be} learning resilient.} Therefore, these tests can only make a concrete decision about whether a scheme is \textit{not resilient}. One iteration of the game is presented in Fig.~\ref{fig:ant}, consisting of the following steps:
\begin{enumerate}
	\item \textbf{T} randomly generates a netlist ($Net$) that consists of any number of gates, where each node can have any number of input or output connections.
	\item \textbf{T} locks the netlist ($Net_{L}$) with a selected scheme using one randomly selected key ($Key$).
	\item \textbf{T} sends the locked netlist to \textbf{U} for analysis.
	\item \textbf{U} guesses the key based on the collected observations.
	\item \textbf{U} sends the guessed key ($Key_{G}$) to \textbf{T} for comparison.
	\item \textbf{T} responds whether the guess is correct or not for each key bit value individually. Note that \textbf{T} does not provide the reason why a bit is correctly or falsely guessed.
\end{enumerate}
Hereby, we assume the following: ($i$) \textbf{T} always uses the same LL scheme and ($ii$) the original netlist is different in each iteration. Hereby, ($i$) ensures that \textbf{U} is able to learn about the LL scheme, otherwise \textbf{T} could always select different schemes. ($ii$) prevents learning from \textit{different key values} for \textit{identical locations} in a netlist. Note that in this game, \textbf{U} does not have to know the exact type of the used LL scheme, just that it is always the same one. The game is repeated until \textbf{U} terminates the game or achieves a desired success rate. 

At the end of an iteration, \textbf{U} is able to update his knowledge with new observations taken from the response provided by \textbf{T}. These observations are used to make guesses about the key in new iterations. If the LL scheme leaks information, \textbf{U} is able to extract meaningful observations leading to the ability to make educated guesses. 
If the guessing accuracy of \textbf{U} continuously increases, then the scheme-under-attack leaks information. If the guessing accuracy remains 50\%, i.e., equal to a random guess, then the collected observations exhibit no evident correlation to the correct key. However, note that this does not prove the absence of leakage, as previously discussed.

Furthermore, it is noteworthy that the only knowledge \textbf{U} has about the netlist is that its original form contains either AND gates exclusively or randomly selected gates. However, \textbf{U} does not know how the gates are initially connected. This is not in line with an actual attack scenario, since an adversary should not be informed about the original gate types of a design. However, this limitation has a favorable connotation; it amplifies the changes induced by an LL scheme. \textit{Therefore, any conclusions made about a scheme in ANT or RNT hold also in a general, for the attacker less invasive case when the adversary is not aware of the original structure of the netlist.}


\subsubsection{ANT} In ANT, \textbf{T} generates netlists consisting exclusively of AND gates, representing the regular end of the spectrum. This feature maximizes the exposure of the inner workings of the LL mechanism, since predictable selections or changes can more easily be spotted in repeating structures. \textit{Evidently, for this particular test, any type of gate would suffice.} In practice, to avoid a potential bias of an LL scheme toward specific gate types, the test should be repeated for all primary types.

\subsubsection{RNT} In RNT, \textbf{T} generates a netlist consisting of a variety of randomly selected gate types. Here, a particular LL scheme might not leak, i.e., reveal information about the key due to the nature of the netlist. For example, the fact that all gate types are present in the netlist might have a favorable implicating on the security of the scheme. If a scheme fails at least one test, it is regarded as conclusively vulnerable, since its resilience depends on the structure of the netlist rather than the key. Moreover, evaluating a scheme through both tests can reveal the cause for information leakage and suggest a specific setting in which resiliency might be achieved.

\subsubsection{Termination Condition} A concrete termination condition cannot be defined in advance. The exact number of iterations depends on the ability of \textbf{U} to extract usable observations, as exemplified in Section~\ref{testing-xor-xnor} and Section~\ref{testing-twin-gate}. 

\subsubsection{Resynthesis and Security}\label{resynthesis-and-security}
Typically, resynthesis is performed after locking is done to further integrate any LL-induced changes. However, as discussed in~\cite{sisejkovic2020challenging}, \textit{an LL scheme must not depend on synthesis to obtain a sufficient security level.} Otherwise, the scheme clearly leaks information if resynthesis is not performed, as its security \textit{depends} on specific synthesis transformations (often part of proprietary software). Moreover, this dependency has an important implication: the security of the system would not solely depend on the key. This can enable novel attacks; for example, the attacker can exploit this dependency by reverse engineering specific synthesis transformations to neutralize the security aspects of a scheme, as performed in the mentioned SAIL attack~\cite{SAIL2019}. To overcome this requirement, the proposed tests \textit{do not include resynthesis}.

\subsubsection{Circuit Family}
Both tests operate on two spectrum ends. In practice, designs might be placed somewhere between these extremes. In this case, a locked netlist might \emph{not} exhibit leakage under suitable structural conditions; implying that the scheme has no identifiable leakage \emph{for this one specific case}. Thus, its security would be predicated on the structural characteristics, i.e., the circuit family, of the underlying netlist. Hence, security conclusions about the challenged scheme cannot be generalized across all netlist structures. Evidently, this must not be the objective for the design of generally secure LL. Therefore, it is beneficial to generalize the evaluation by looking at the spectrum ends. Nevertheless, designing LL that is specific to a circuit family is a valid objective. In this context, both ANT and RNT can help uncover \emph{for which circuit family the evaluated scheme might exhibit resilience}. This allows for the design of LL schemes which are structurally secure under the assumption of a certain circuit family; which can be a reasonable use case in practice. Examples of such cases are presented in Section~\ref{testing-xor-xnor} and~\ref{testing-twin-gate}.



\begin{figure}[t]
	\centering
	\subfloat[ANT]{
		\includegraphics[width=0.42\columnwidth]{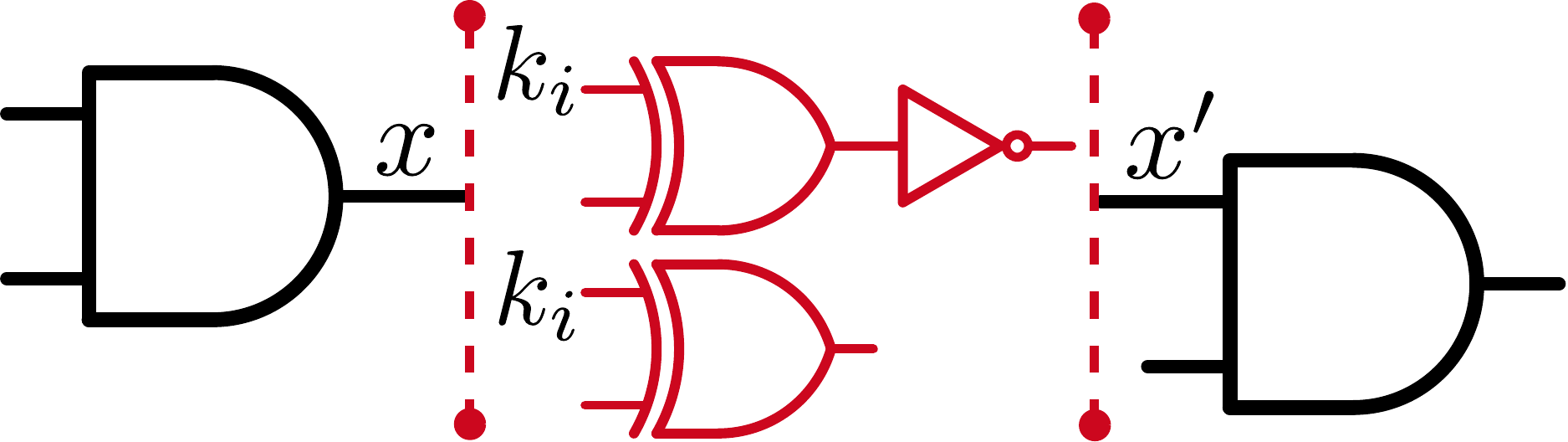}
	}
	\subfloat[RNT]{
		\includegraphics[width=0.42\columnwidth]{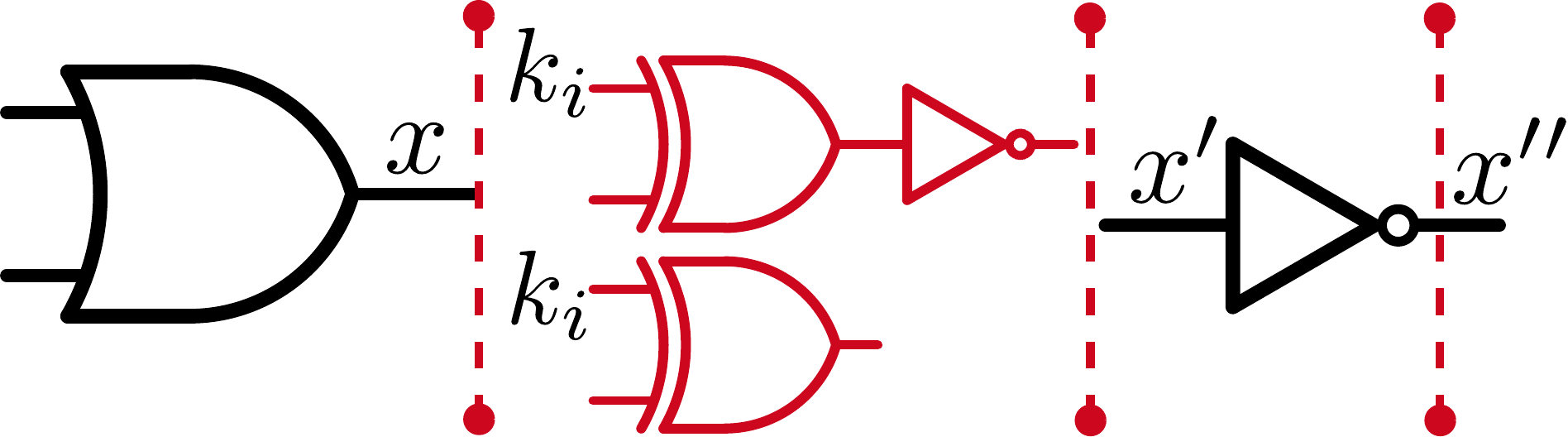}
	}
	\caption{ANT and RNT for XOR/XNOR-based locking.}
	\label{fig:test-epic}
\end{figure}
\subsection{Testing the XOR/XNOR Scheme}\label{testing-xor-xnor}
Let us first consider EPIC as a representative of XOR/XNOR-based locking.
In EPIC, the location selection is random.
 However, EPIC has been successfully attacked with ML-based attacks~\cite{SAIL2019}. Therefore, EPIC is leaking information through the introduced change, as argued in~\cite{sisejkovic2020challenging}. 

\subsubsection{ANT Observations}
In ANT, \textbf{T} locks netlists that consist exclusively of AND gates. After many iterations, \textbf{U} is able to learn that an XOR implies 0, while XNOR implies 1. This can be understood through the visualization in Fig.~\ref{fig:test-epic}~(a). In ANT, \textbf{U} is aware that the original netlist exclusively contains AND gates. Thus, it is possible for \textbf{U} to isolate the \textit{exact} location that is affected by an XOR or XNOR between two AND gates. In the given example, it is evident that the value $x$ entering a key gate must be preserved once processed. Therefore, the adversary has to guess the value of $k_{i}$ that ensures this preservation, i.e., $k_{i}=? \Rightarrow{x=x'}$. Due to the nature of the scheme and ANT, it is always true that $k_{i}=0$ for XOR and $k_{i}=1$ for XNOR. 

\textbf{Verdict:}~\textbf{U} is able to easily learn about the key by isolating the limited region around a key gate. \textit{Therefore, EPIC fails ANT and is considered vulnerable to learning-based attacks.}

\subsubsection{RNT Observations}
In RNT, \textbf{T} generates a netlist using a variety of gate types for each iteration. Thus, the possibility exists that an XOR key gate is placed before an original inverter. This leads to the occurrence of observations where an XOR+INV is associated both with the key value 0 and the value 1. In other words, sometimes it is more difficult for \textbf{U} to correctly isolate the key gates, as shown in Fig.~\ref{fig:test-epic}~(b). Here, it is not clear where the correct cut-off line for the restoration of $x$ is drawn, i.e., whether $x=x'$ or $x=x''$. Therefore \textbf{U} is more likely to make \textit{false guesses} when these situations occur. 

\textbf{Verdict:}~In RNT, \textbf{U} can sometimes make a false guess. However, since \textbf{T} generates netlists with an evenly distributed number of gate types, these contradicting cases do not occur often. Consequently, it is expected that the guessing accuracy in RNT is higher than 50\%. Therefore, \textit{EPIC fails RNT.}
\begin{figure}[t]
	\centering
	\subfloat[Case 1]{
		\includegraphics[width=0.49\columnwidth]{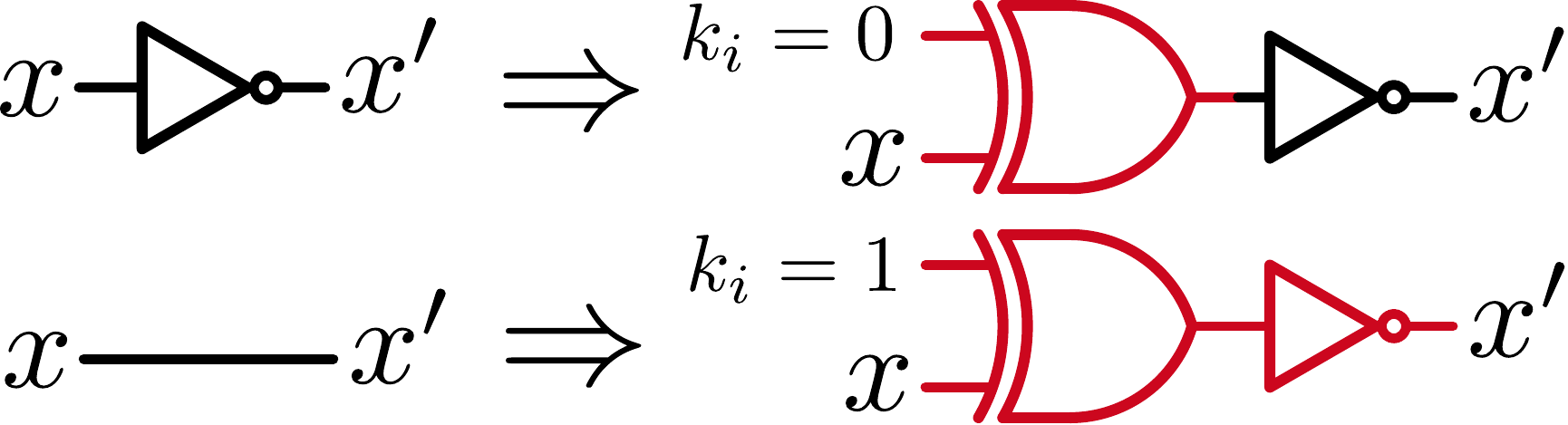}
	}
	\subfloat[Case 2]{
		\includegraphics[width=0.42\columnwidth]{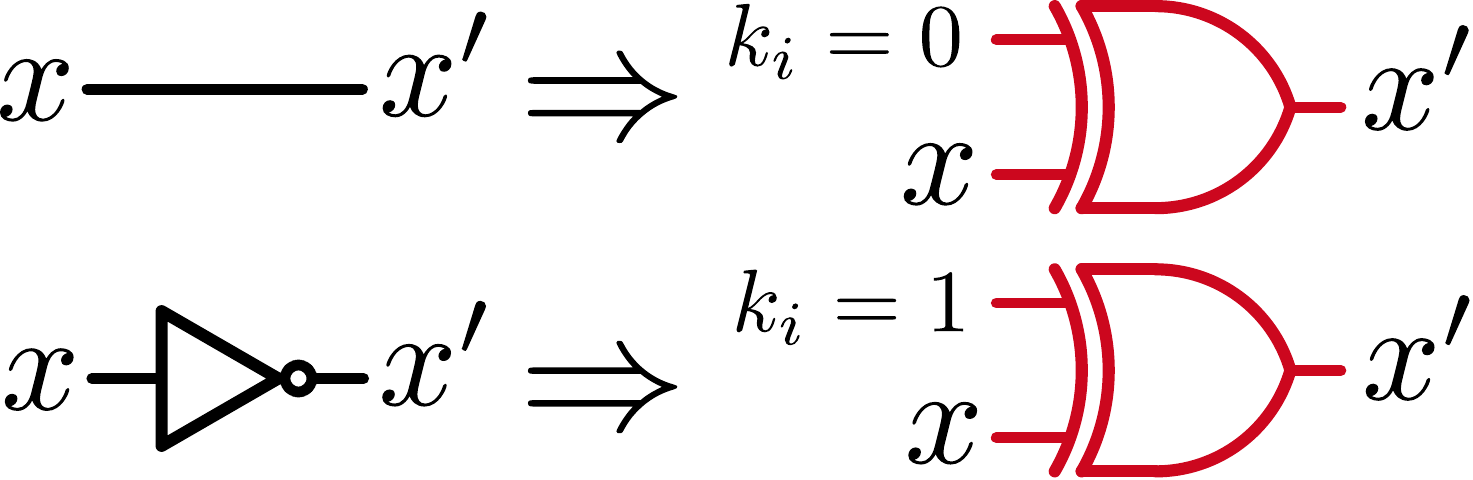}
	}
	\caption{XOR/XNOR learning resilience cases.}
	\label{fig:test-epic-special-cases}
\end{figure}

\textbf{Discussion:}~
Based on the tests, we can conclude that XOR/XNOR LL might exhibit learning resilience only in \textit{very specific} cases, including: XOR key gates are only placed in front of existing inverters (Fig.~\ref{fig:test-epic-special-cases}~(a)) or XOR gates replace existing inverters for key bit 1, i.e., XNOR gates are not used (Fig.~\ref{fig:test-epic-special-cases}~(b)). This concept has been implemented in Truly Random LL (TRLL)~\cite{TRLL2020}. Even though both cases lead to contradicting observations, the learning resilience is only enabled by the \textit{inherent structural features} of the design. Therefore, TRLL fails ANT since no inverters are available to be replaced or coupled with an XOR gate; disintegrating TRLL into traditional XOR/XNOR LL. However, TRLL passes RNT since a sufficient amount of gate types is available to support the scheme. These examples also suggest how to achieve learning resilience; \textit{equivalent local regions are labeled with the same key-bit}. Here, the attacker has a 50\% chance to guess correctly. Nevertheless, it still remains a challenge to support this concept and satisfy both tests using XOR/XNOR gates.




\subsection{Testing the Twin-Gate Scheme} \label{testing-twin-gate}
To present other aspects of the tests, we propose a new LL scheme dubbed \textit{Twin-Gate} (Algorithm~\ref{alg:twin-gate}). Twin-Gate receives the correct key and a netlist as input. First, Twin-Gate prepares the set of all gate types with more than one input ($T_{multi}$, line 1) and the set of all single-input gate types ($T_{single}$, line 1). Afterwards, for each key bit (line 2), Twin-Gate performs three steps: the random selection of a \textit{true} node (line 3), the creation of a \textit{false} node (lines 5-11), and the assembly of a replacement MUX (lines 13-14). The true node represents an original node from the netlist. The false node is the additional \textit{twin} that is added as a pair to the true node. Depending on the type of the selected true node, the algorithm selects a suitable false node. The suitability is defined by ensuring that both nodes have the same number of inputs and a different type. Therefore, if the true node is of type INV or BUF, its twin can only be BUF or INV (line 6). In case the true node is a multi-input node, the algorithm randomly selects a type from the available gate types (line 8). Once the false node is created (line 11), a MUX is assembled by coupling its inputs to the true and false node outputs (line 13). The current key input ($K[i]$) acts as selection bit and determines which gate output is forwarded. A locked example is shown in Fig.~\ref{fig:twin-gate}~(a) based on the original netlist shown in Fig.~\ref{fig:enc-example}~(a). Here, the true node $G_{3}$ is replaced by the pair $(KG_{1}, KG_{2})$. The correct gate output is selected by the multiplexer through the key bit $k_{1}$.


\subsubsection{ANT Observations}
\begin{figure}[b]
	\centering
	\subfloat[Locked example]{
		\includegraphics[width=0.37\columnwidth]{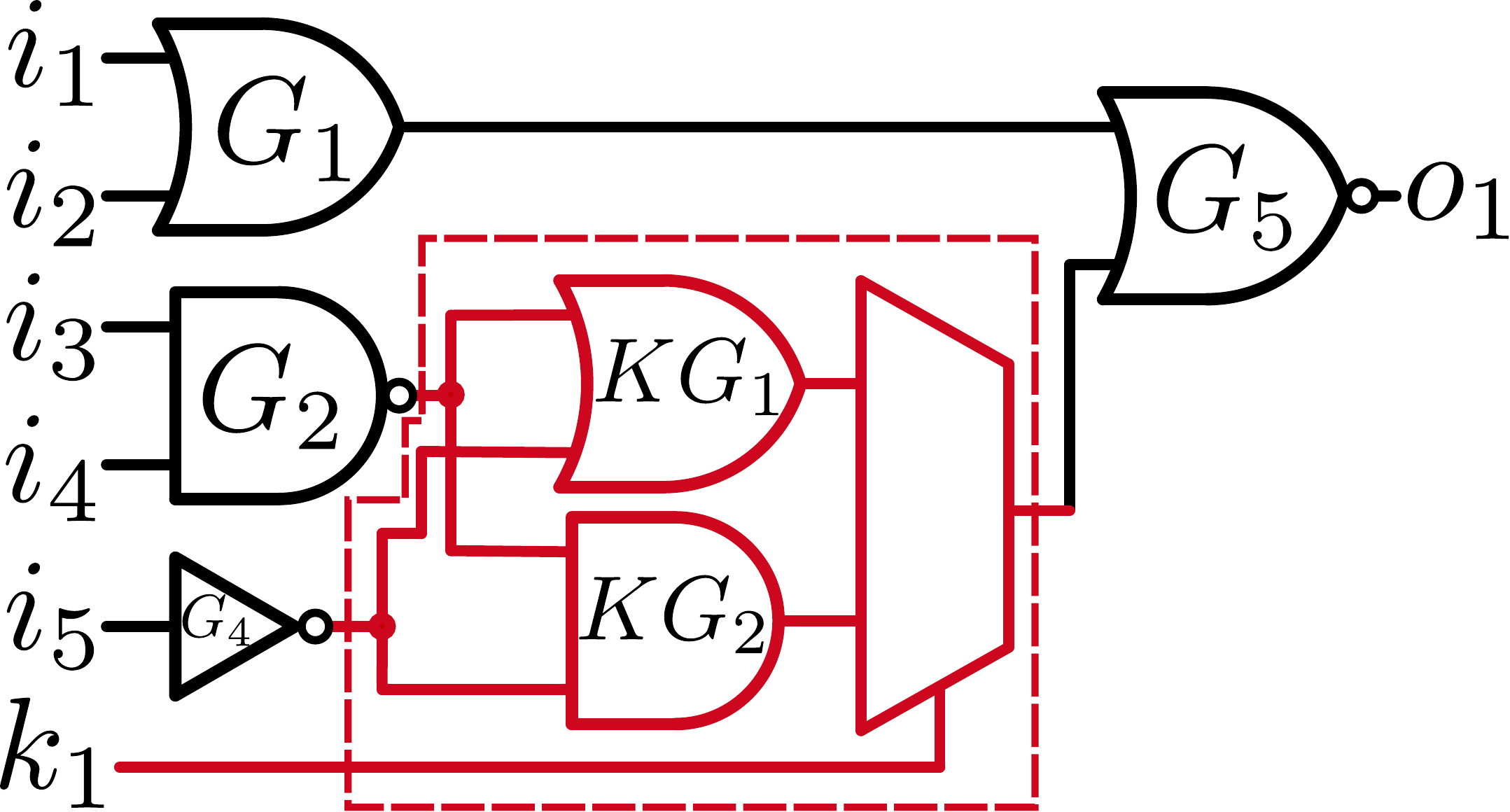}
	}
	\subfloat[ANT analysis]{
		\includegraphics[width=0.37\columnwidth]{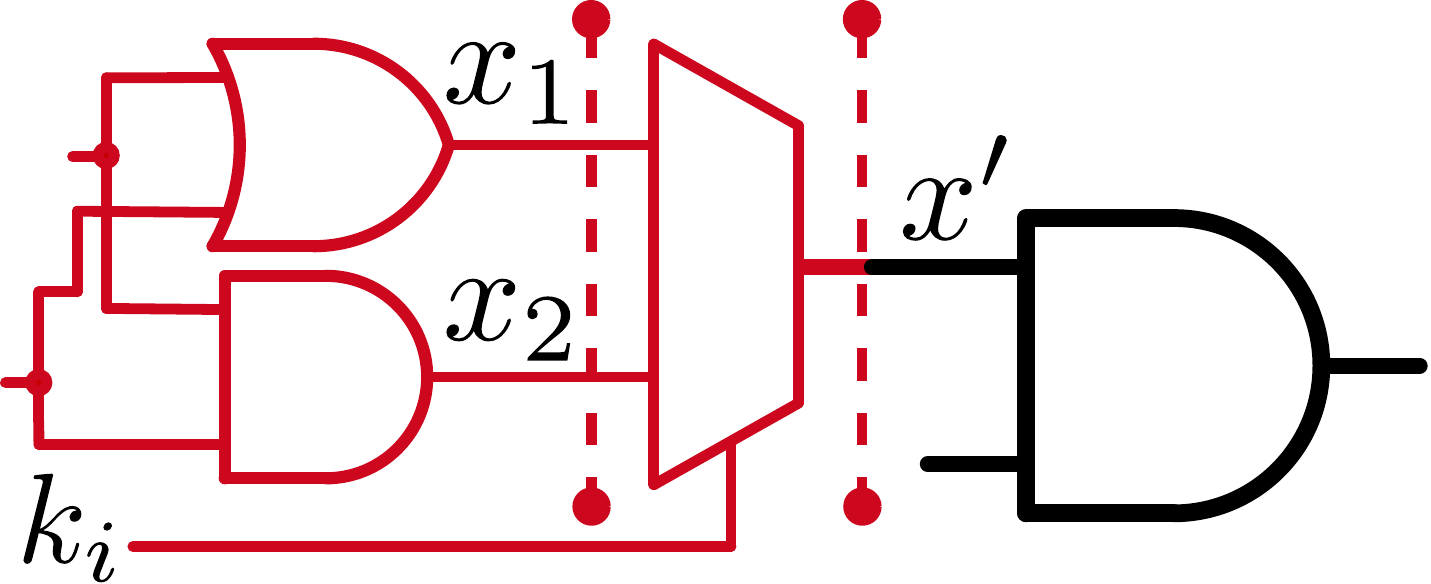}
	}
	\caption{Twin-Gate locking scheme.}
	\label{fig:twin-gate}
\end{figure}

In ANT, Twin-Gate resolves the \textit{direct leakage} problem of XOR/XNOR-based locking, as now \textbf{U} is not able to guess anything about the nature of the key based on the induced gate types if only considering the local change. Let us consider the example in Fig.~\ref{fig:twin-gate}~(b). Here, Twin-Gate replaces an original AND gate with the pair (OR, AND). Due to the nature of Twin-Gate, it is possible to isolate which gates are part of the LL scheme. In this case, identifying the correct key $k_{i}$ is equivalent to correctly guessing whether $x'=x_{1}$ or $x'=x_{2}$.  Here the adversary is not able to guess whether AND or OR is the true node by analyzing the samples individually, since both are valid options. However, it turns out that \textbf{U} is able to learn about the key based on the type and distribution of \textit{all} key gates in the netlist. For example, after many iterations of the game, \textbf{U} can learn that AND is \textit{never} a false gate, as otherwise two ANDs are fed to a MUX; implying that any key bit value is correct. Since AND is never a false gate, \textbf{U} can easily learn which key bit is correct for a given pair. This effect is amplified by the nature of AND-based netlists that exclude AND gates as possible false gates. 
%

\textbf{Verdict:}~Twin-Gate fails ANT as the selection of false nodes directly depends on the original netlist structure. However, RNT uncovers specific cases where the scheme might be learning resilient, as discussed in the following. 
%
%
\begin{algorithm}[t!]\scriptsize
	\caption{Twin-Gate locking scheme}\label{alg:twin-gate}
	\begin{algorithmic}[1]
		\Require{Activation key $K$, netlist $Net$}
		\Ensure{Locked netlist}
		\State $T_{multi}$ $\gets \{$AND, OR, $\dots$, XOR$\};~T_{single} \gets \{$NOT, BUF$\}$
		\For{$i = 0$ to $\abs{K}$}
		\State $n_{true}$ $\gets$ RandomSelection($Net$)
		\LineComment{Create false node}
		\If{TypeOf($n_{sel}$) $\in~\{$NOT, BUF$\}$}
		\State $F_{types} \gets \{t\in{T_{single}}~|~t\neq~$TypeOf($n_{true}$)$\}$
		\Else
		\State $F_{types} \gets \{t\in{T_{multi}}~|~t\neq~$TypeOf($n_{true}$)$\}$
		\EndIf
		\State	$f_{type} \leftarrow$ RandomSelection($F_{types}$)
		\State	$n_{false} \leftarrow$ CreateNodeOfType($f_{type}$)
		\LineComment{Assemble MUX node}
		\State	$m_{enc} \leftarrow $CreateMUXFor($n_{true}, n_{false}, K[i]$)
		\State	ReplaceNodes($n_{true}, m_{enc}$)
		\EndFor
		\State \textbf{return} $Net$
	\end{algorithmic}
	
\end{algorithm}

\subsubsection{RNT Observations}
Since each iteration of RNT is based on a netlist with randomly selected gate types, \textbf{U} has a disadvantage in guessing the correct key value. Now, every gate type is equally likely to be selected as true or false node, making it difficult to extract meaningful observations.

\textbf{Verdict:}~In theory, Twin-Gate passes RNT. However, in practice, typical circuits do not have a perfectly balanced amount of each gate type, making it possible to identify potential false nodes. Moreover, the presented discussion showcases how even a global leakage can be identified in LL schemes using the proposed tests. Hereby, global refers to the leakage being uncovered due to a frequency analysis over all gates.

\subsection{Lessons Learned}\label{lessons-learned}

Based on the analysis, we can conclude that a learning-resilient scheme must have the following properties: $(i)$ its security must not depend on resynthesis, $(ii)$ the induced change must not depend on the inherent structural features of the original netlist, and $(iii)$ the induced change and its location should not depend on the value of the connected key bit. Hereby, $(iii)$ is manifested in the following; the observations made by \textbf{U} must be either random (noisy data) or two identical observations must suggest two different key values (e.g., two identical "images" are labeled differently). For example, XOR/XNOR-based locking is 
vulnerable because, in most cases, it cannot fulfill the latter.

Testing whether a human adversary (or, ultimately, an ML system) is able to challenge a scheme through \textit{learning-based attacks} is a fundamentally important step in designing LL, as it tests the basic security foundation; \emph{can the adversary guess the key based on the induced change?} Therefore, the introduced theoretical tests offer a simple yet powerful approach to uncover elemental security flaws in LL, \textit{even before an implementation or empirical results are available.}


\begin{figure}[b]
	\centering
	\subfloat[MUX locking]{
		\includegraphics[width=0.33\columnwidth]{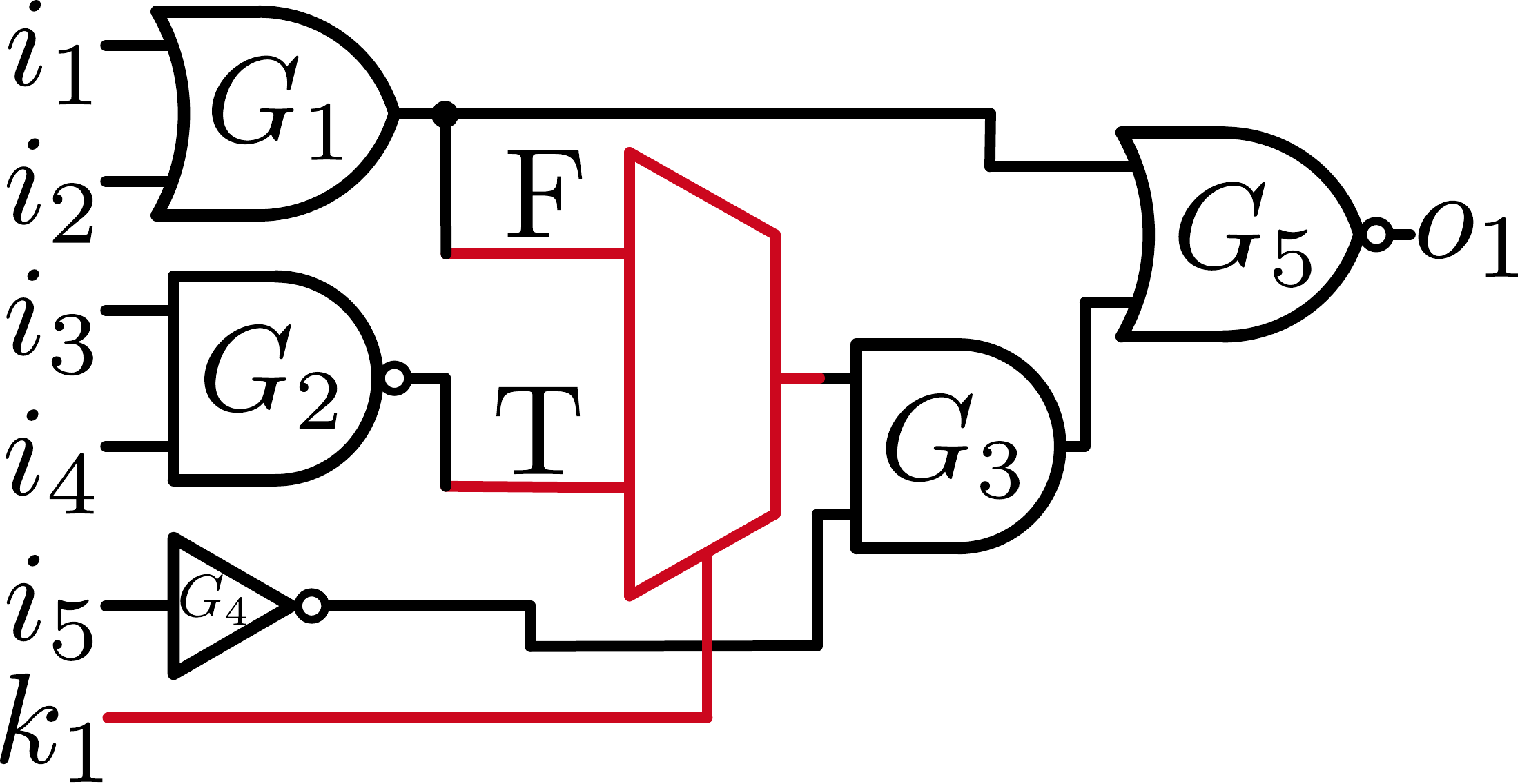}
	}
	\subfloat[True wire selected]{
		\includegraphics[width=0.3\columnwidth]{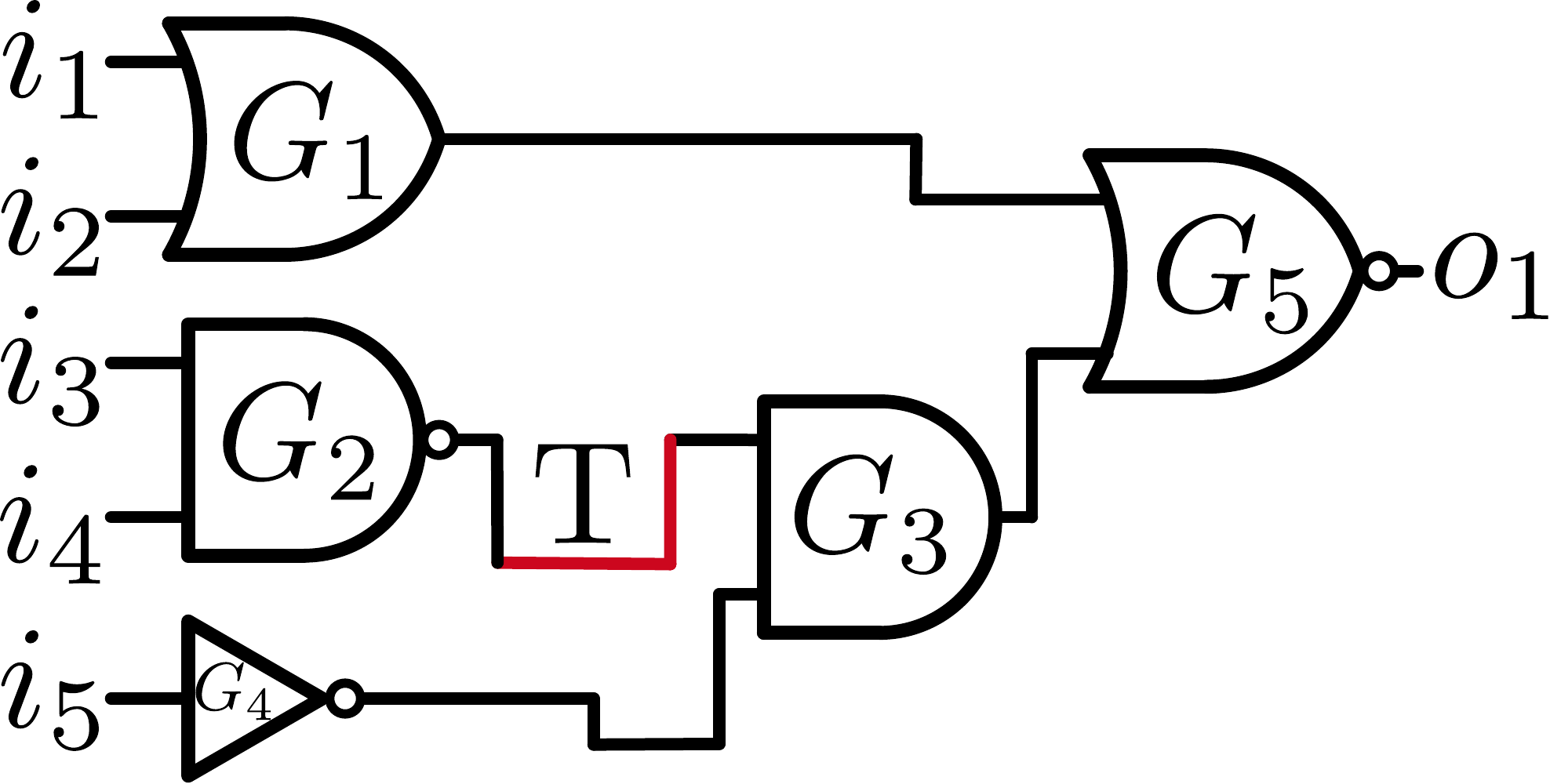}
	}
	\subfloat[False wire selected]{
		\includegraphics[width=0.3\columnwidth]{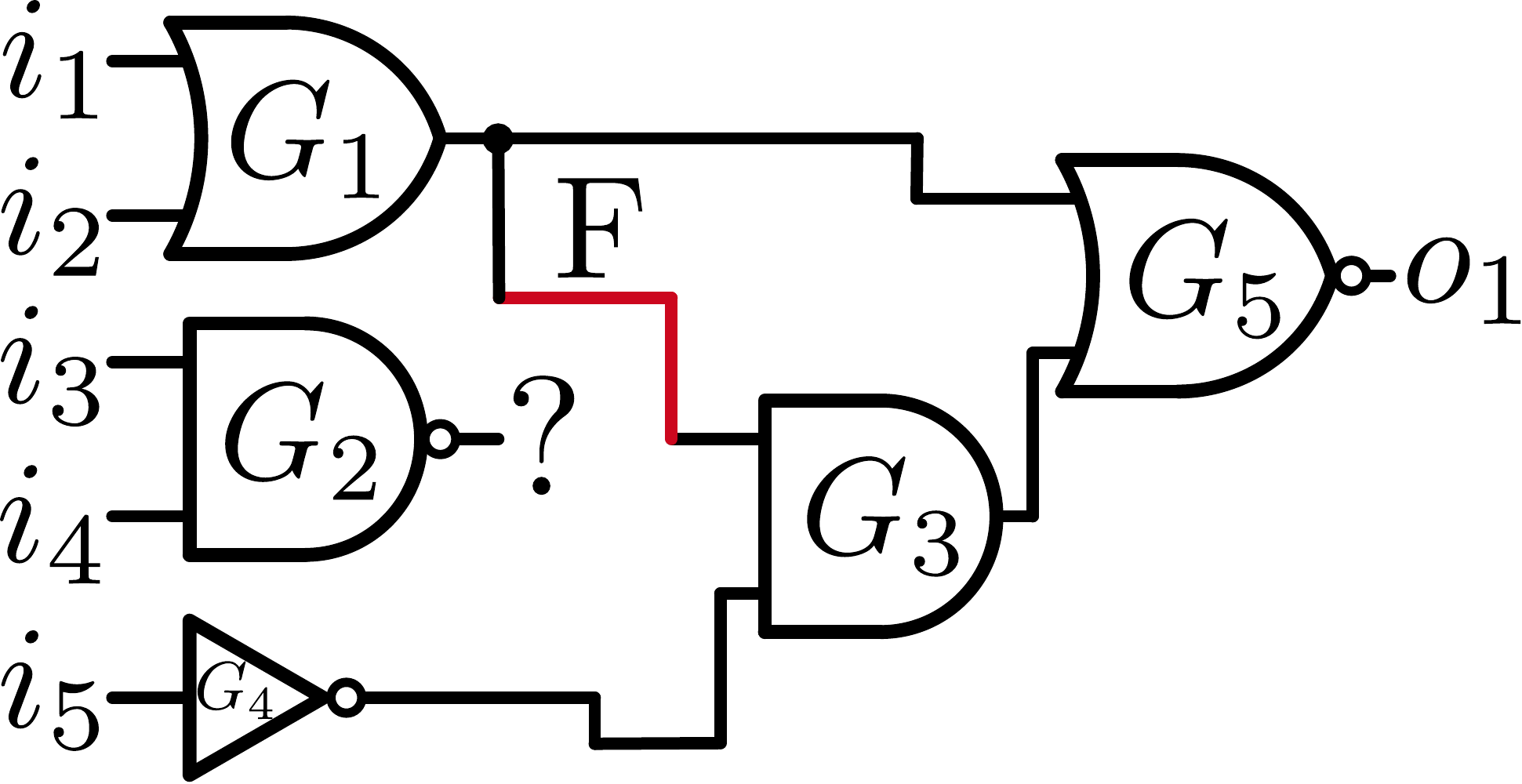}
	}
	\caption{Example: MUX-based locking and SAAM.}
	\label{fig:saam-example}
\end{figure}

\section{SAAM: Structural Analysis Attack on MUX-Based Logic Locking}\label{saam:attack}
Through ANT and RNT, we identified an important vulnerability of MUX-based LL that can determine the correct key bit value for a single MUX. The MUX takes two inputs: the true (T) and false (F) wire. Both values of the key bit result in a functionally valid netlist. 
However, by analyzing both gates that drive T and F, we can easily identify the correct true wire as follows. Let us consider the example shown in Fig.~\ref{fig:saam-example}~(a). Here, the original netlist from Fig.~\ref{fig:enc-example} is locked with a MUX. In this setting, the adversary can create two cases: for each value of the selecting key bit, the MUX is removed and the selected wire is forwarded. If the MUX selects T as the correct wire (Fig.~\ref{fig:saam-example}~(b)), the resulting netlist displays no structural faults. However, if the MUX selects F as the correct wire (Fig.~\ref{fig:saam-example}~(c)), the wire T (output of gate G2) remains an \textit{unconnected} (dangling) wire. Moreover, this is \textit{never} the case for the wire F, as it is randomly selected out of the other wires in the netlist. Thereby, the gate output that drives F (gate G1) is \textit{always} connected to some other gate as well (gate G5). \textit{Therefore, the MUX input that remains unconnected when not selected is the true wire.} To the best of our knowledge, this major design fallacy in MUX-based locking has not been documented properly yet; therefore, we summarize it in the form of a simple attack named SAAM; \textbf{S}tructural \textbf{A}nalysis \textbf{A}ttack on \textbf{M}UX-based locking. SAAM is presented in Algorithm~\ref{SAAM}. SAAM iterates through each key input and checks which input of the connected MUX has one output (line 4 to 10). Note that SAAM fails at uncovering the key value if both inputs result in connected logic if not selected. This can also occur in the existing MUX-based LL if, by chance, a favorable T wire is selected.

\begin{algorithm}[t] \scriptsize
	\caption{SAAM: \textbf{S}tructural \textbf{A}nalysis \textbf{A}ttack on \textbf{M}UX-based locking}
	\label{SAAM}
	\begin{algorithmic}[1]
		\Require{Locked netlist $Net$}
		\Ensure{Extracted key $E$ where $\forall{e}\in{\{0, 1, X\}}$}
		\State $E \gets \{\emptyset\}$
		\For{$i = 0$ to $\abs{K}$}
		\State $\{n_{1}, n_{2}\} \gets$ ExtractDirectInputsOfMUX($Net$, $K[i]$)
		\If{(OutSize($n_{1}$) \textgreater ~1)$~\land~$(OutSize($n_{2}$) \textgreater ~1)}
		\State $E[i] \gets{X}$\Comment{Set unknown value}
		\ElsIf{OutSize($n_{1}$) == 1}
		\State $E[i] \gets$ ValueOfKeyForMUXInput($n_{1}$)\Comment{Out of $n_{1}$ is true wire}
		\Else
		\State $E[i] \gets$ ValueOfKeyForMUXInput($n_{2}$)\Comment{Out of $n_{2}$ is true wire}
		\EndIf
		\EndFor
		\State \textbf{return} $E$
	\end{algorithmic}
\end{algorithm}
Interestingly, previous works showcase MUX-based LL including similar examples without noticing this vulnerability~\cite{plaza2015solving,logicConeAnalysis2015, fault2015}. The authors in~\cite{sweepAttack2019} briefly mention that this might be the reason why SWEEP is effective against MUX-based LL. 

\section{Designing a Deceptive Logic-Locking Scheme}\label{dmux:description}
The reason for the vulnerability against learning-based attacks lies in the challenge of inserting \textit{additional logic} without leaving key-related, structural traces. A promising LL scheme that overcomes these issues hides in MUX-based locking. MUX-based LL has a profound advantage; instead of inserting \textit{additional gates}, it reconfigures the \textit{existing logic}. Thereby, the scheme always inserts the same structures: multiplexers. However, we have shown that existing MUX-based LL can be attacked efficiently with SAAM and SWEEP. Therefore, based on MUXs, we introduce D-MUX; a deceptive LL scheme that overcomes both attacks and offers effective learning resilience. The core functionality of D-MUX is based on specific \textit{locking strategies} ensuring that each path through a MUX has the same probability of being true or false.
In the following, we first provide more details on the concept of locking strategies. Afterwards, using these strategies, we compose D-MUX.

\subsection{Locking Strategies}
\begin{figure}[t]
	\vspace{-0.1in}
	\centering
	\subfloat[$S_{1}$]{
		\includegraphics[valign=c, width=0.2\columnwidth]{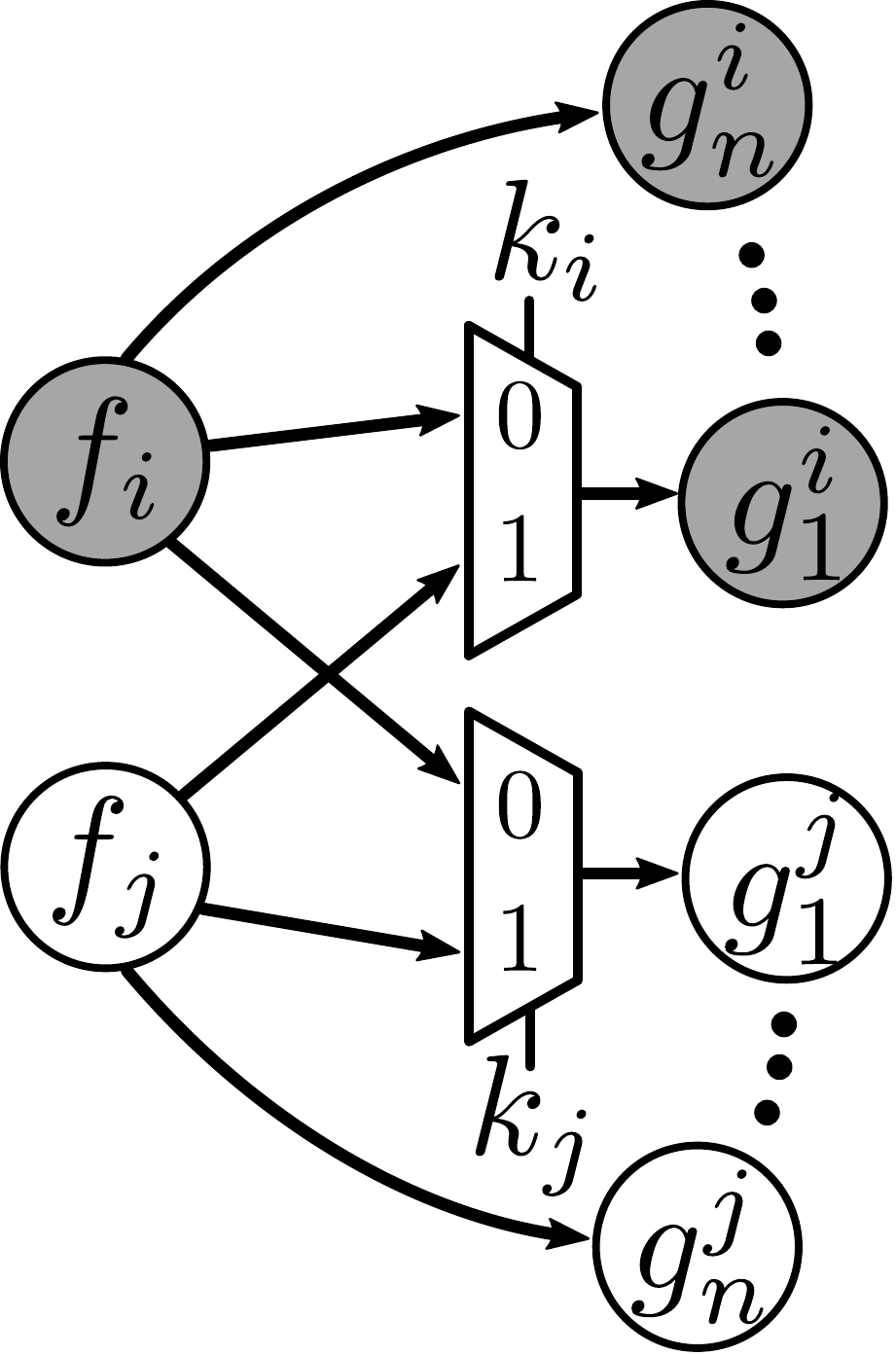}}\hfil
	\subfloat[$S_{2}$]{
		\includegraphics[valign=c,width=0.2\columnwidth]{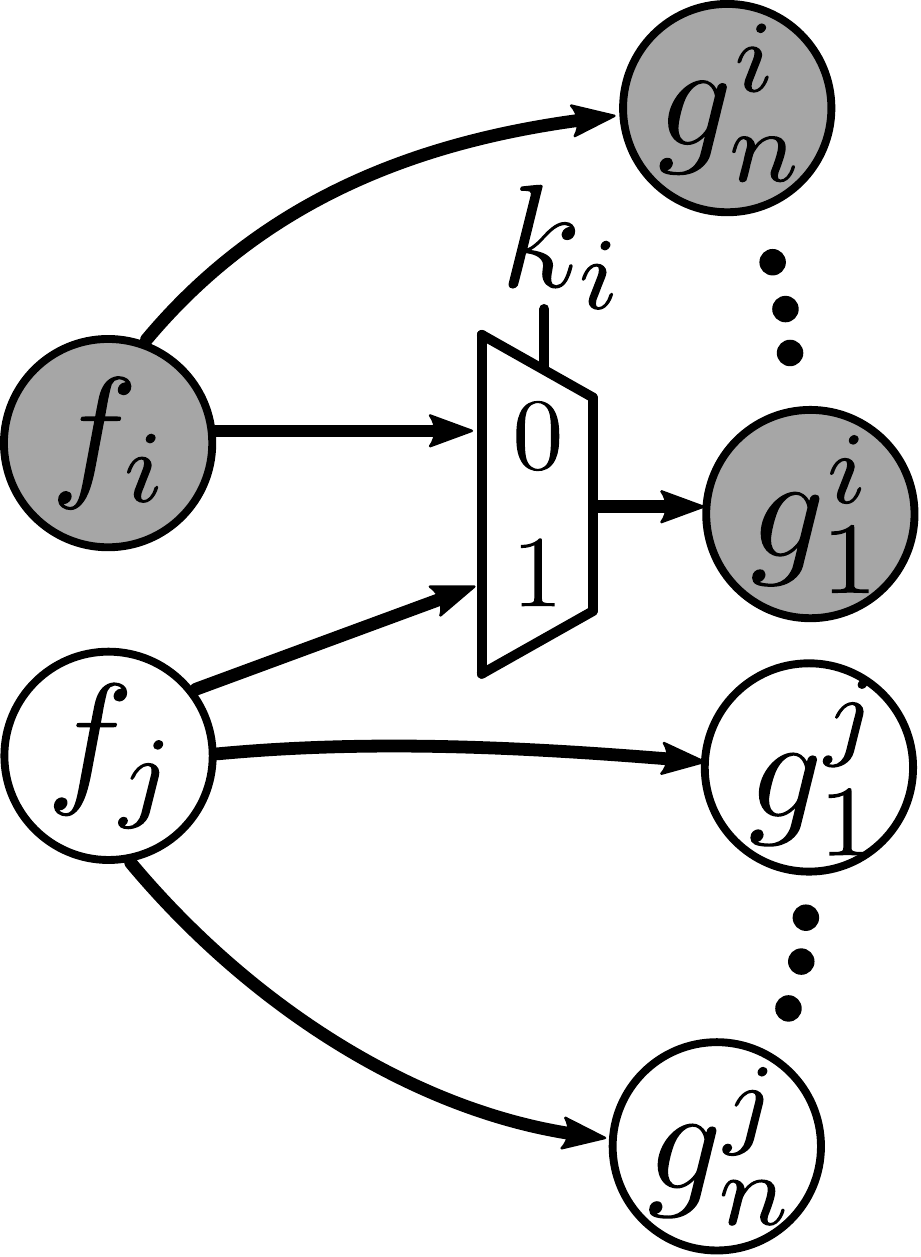}
		\vphantom{\includegraphics[width=0.11\columnwidth,valign=t]{example-image-10x16}}%
	}\hfil
	\subfloat[$S_{3}$]{
		\includegraphics[valign=c,width=0.2\columnwidth]{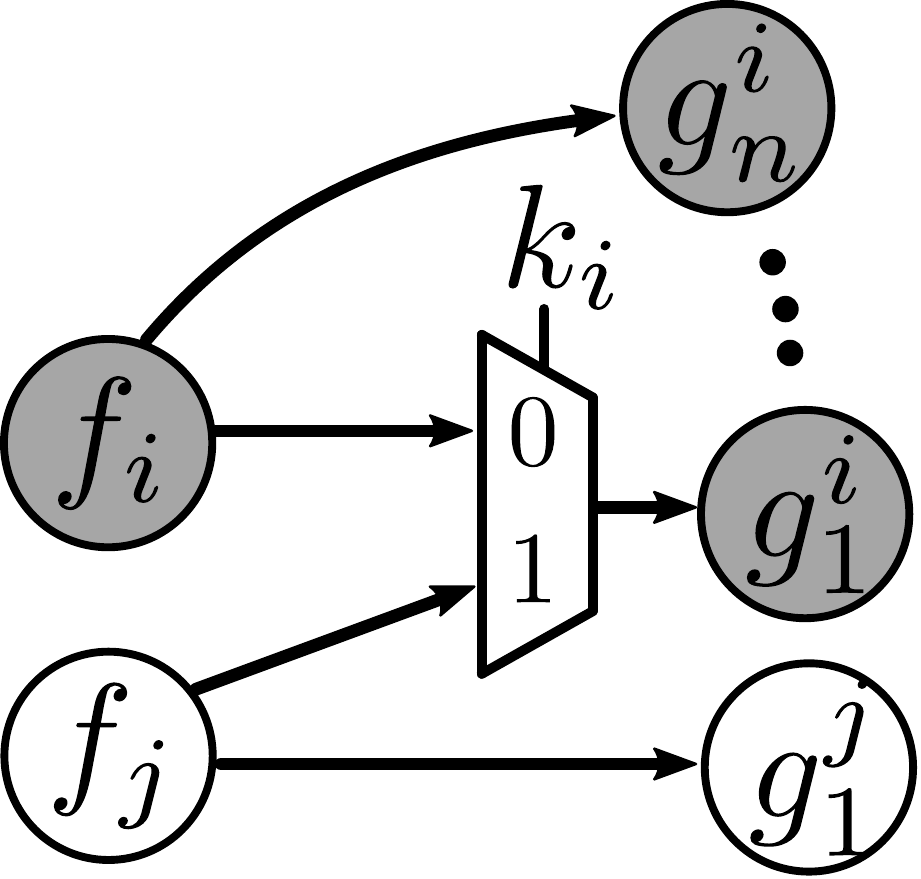}
		\vphantom{\includegraphics[width=0.11\columnwidth,valign=t]{example-image-10x16}}%
	}\hfil
	\subfloat[$S_{4}$]{
		\includegraphics[valign=c,width=0.2\columnwidth]{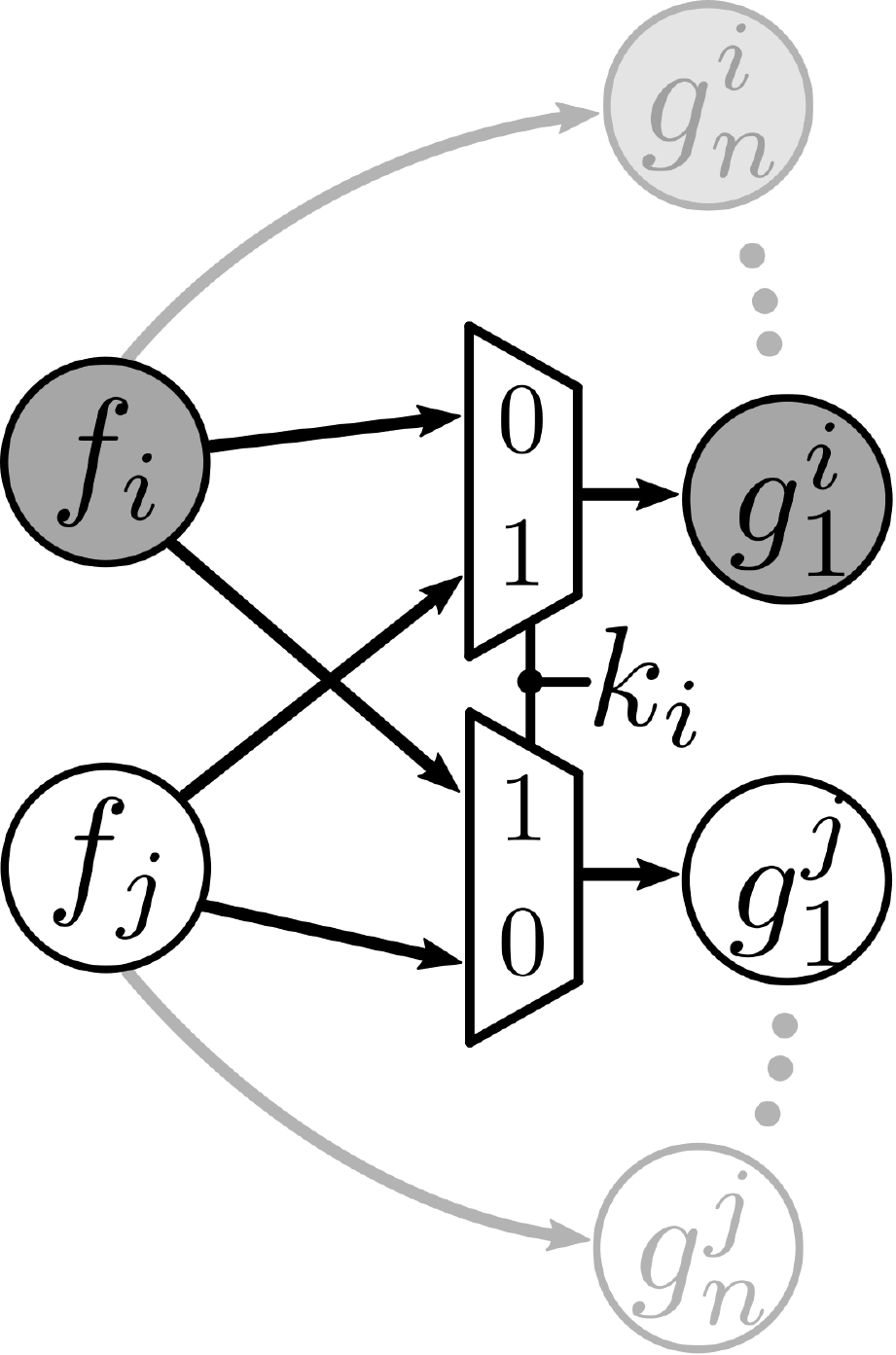}
		\vphantom{\includegraphics[width=0.11\columnwidth,valign=t]{example-image-10x16}}%
	}	
	\caption{D-MUX locking strategies.}
	\label{fig:strategies}
\end{figure}

To achieve resilience against SAAM, all inputs to a MUX must be equally likely. To dissolves the possibility of an educated guess determining the correct (true) wire, the wire selection and MUX insertion must be steered to avoid selecting single output gates as candidates. Therefore, we introduce multiple locking strategies that fulfill this criterion. A single locking strategy $S_{i}$ is defined by four components:

\begin{itemize}
	\item \textbf{Input Node Selection:} Selects a set of two input nodes $\{f_{i}, f_{j}\}$. These nodes represent two gates that drive the inputs of one or multiple MUXs.
	\item \textbf{MUX Configuration Selection:} Selects a set of maximum two MUXs to be used in a single locking iteration. 
	\item \textbf{Key Length Selection:} Selects a set of maximum two one-bit key inputs $\{k_{i}, k_{j}\}$.
	\item \textbf{Output Node Selection:} Selects a set of maximum two output nodes $\{g^{i}, g^{j}\}$. Hereby, $g^{i}$ is the output node of $f_{i}$ if $f_{i}$ drives one input of $g^{i}$ in the target netlist. 
\end{itemize}
The node $f_{i}$ can be of the following types:
\begin{itemize}
	\item \textbf{Single-output:}  $f_{i}$ has only one output node $g_{1}^{i}$.
	\item \textbf{Multi-output:} $f_{i}$ has multiple output nodes $\{g_{1}^{i},g_{2}^{i},\dots\}$.
\end{itemize}
Based on these components, several locking strategies can be derived:
\textit{TwoMultiOutTwoBitTwoMux} ($S_{1}$), \textit{TwoMultiOutOneBitOneMux} ($S_{2}$), \textit{OneMultiOutOneBitOneMux} ($S_{3}$), and \textit{AnyOutOneBitTwoMux} ($S_{4}$). In the following, all strategies are discussed in detail, following the examples in Fig.~\ref{fig:strategies}.
\begin{table}[b]
	\caption{Strategy configurations.}
	\centering
	\subfloat[$S_{1}$]{	\tabcolsep=0.11cm
		\begin{tabular}{c|c?c|c}
			\bm{$k_{i}$} & 	\bm{$k_{j}$} & 	\bm{$f_{i}$} & 	\bm{$f_{j}$} \\ \Xhline{2\arrayrulewidth}
			$0$ & $0$ & $g_{1}^{i},g_{1}^{j}$ & $\emptyset$  \\ \hline
			$0$ & $1$ & $g_{1}^{i}$ & $g_{1}^{j}$  \\ \hline
			$1$ & $0$ & $g_{1}^{j}$ & $g_{1}^{i}$  \\ \hline
			$1$ & $1$ & $\emptyset$ & $g_{1}^{i},g_{1}^{j}$  \\
		\end{tabular}	
	}\hfil\subfloat[$S_{2}$]{	\tabcolsep=0.11cm
		\begin{tabular}{c?c|c}
			\bm{$k_{i}$} & 	\bm{$f_{i}$} & 	\bm{$f_{j}$} \\ \Xhline{2\arrayrulewidth}
			$0$ & $g_{1}^{i}$ & $\emptyset$  \\ \hline
			$1$ & $\emptyset$ & $g_{1}^{i}$  \\
	\end{tabular}}\hfil\subfloat[$S_{3}$]{	\tabcolsep=0.11cm
		\begin{tabular}{c?c|c}
			\bm{$k_{i}$} & 	\bm{$f_{i}$} & 	\bm{$f_{j}$} \\ \Xhline{2\arrayrulewidth}
			$0$ & $g_{1}^{i}$ & $\emptyset$  \\ \hline
			$1$ & $\emptyset$ & $g_{1}^{i}$  \\
	\end{tabular}}\hfil\subfloat[$S_{4}$]{	\tabcolsep=0.11cm
		\begin{tabular}{c?c|c}
			\bm{$k_{i}$} & 	\bm{$f_{i}$} & 	\bm{$f_{j}$} \\ \Xhline{2\arrayrulewidth}
			$0$ & $g_{1}^{i}$ & $g_{1}^{j}$  \\ \hline
			$1$ & $g_{1}^{j}$ & $g_{1}^{i}$  \\
	\end{tabular}}
	\label{tab:locking:configurations}
\end{table}
\subsubsection{$S_{1}$ - TwoMultiOutTwoBitTwoMux} This strategy selects two multi-output nodes $\{f_{i}, f_{j}\}$ (TwoMultiOut), thereby introducing a pairwise lock by using two individual key bits $\{k_{i}, k_{j}\}$ (TwoBit). Every key bit acts as selector for one particular MUX (TwoMux). A visualization is presented in Fig.~\ref{fig:strategies}~(a). One output node is selected for each input node, i.e., $\{g^{i}, g^{j}\}$, to select two initial \textit{true paths}: $f_{i}\rightarrow{g_{1}^{i}}$ and $f_{j}\rightarrow{g_{1}^{j}}$. Afterwards, two MUXs are placed between the two input and two output nodes to generate four valid paths for all values of $\{k_{i}, k_{j}\}$. Since both input nodes initially have multiple outputs, one cannot determine which path is true or false, as all are equally valid. Moreover, a path from an input to an output node does not have to exists at all. For example, if $k_{i} = 0$ and $k_{j} = 0$, input node $f_{j}$ is neither driving $g_{1}^{i}$ nor $g_{1}^{j}$. This is, however, valid, since $f_{j}$ is a multi-output node and therefore remains connected even if not selected by any MUX. All allowed configurations of $S_{1}$ are shown in Table~\ref{tab:locking:configurations}~(a). The entries are read as follows. For the input keys $k_{i}$ and $k_{j}$, the nodes $f_{i}$ and $f_{j}$ are forwarded to the nodes marked in the table. For example, if $k_{i} = 0$ and $k_{j} = 1$, the output of $f_{i}$ is connected to $g_{1}^{i}$, whereas $f_{j}$ is connected to $g_{1}^{j}$.

\subsubsection{$S_{2}$ - TwoMultiOutOneBitOneMux} This strategy selects two multi-output nodes $\{f_{i}, f_{j}\}$ (TwoMultiOut), but introduces a locking mechanism based on a single key bit $k_{i}$ (OneBit) driving a single MUX (OneMux). In the next step, $S_{2}$ randomly selects one output node of a randomly selected input node. For instance, in the example in Fig.~\ref{fig:strategies}~(b), $S_{2}$ selects $f_{i}$ and one of its output nodes $g^{i}_{1}$. The MUX is placed between these two nodes, enabling the two configurations presented in Table~\ref{tab:locking:configurations}~(b).
Since both input nodes originally have multiple outputs, the case when one of the input nodes is not forwarded to an output via the MUX remains valid.

\subsubsection{$S_{3}$ - OneMultiOutOneBitOneMux}
In this strategy, only one multi-output node ${f_{i}}$ is selected (OneMultiOut) to be locked using a single key bit $k_{i}$ (OneBit) driving one MUX (OneMux), as presented in Fig.~\ref{fig:strategies}~(c). $S_{3}$ enables two configurations as shown in Table~\ref{tab:locking:configurations}~(c). Even though these configurations are, in principle, equivalent to the previous strategy, $S_{3}$ differs from $S_{2}$ only in the fact that the output node must be selected from the multi-output input node. Hereby, regardless of which key bit is chosen, both input nodes remain connected.
\subsubsection{$S_{4}$ - AnyOutOneBitTwoMux} 
This strategy selects two input nodes $\{f_{i}, f_{j}\}$  from the set of all available nodes (AnyOut). Afterwards, a single key bit $k_{i}$ is selected (OneBit) to drive two MUXs simultaneously (TwoMux). For each input node, one output node is selected. The MUXs are configured to always forward the results of both input gates (otherwise SAAM is applicable in case a single-output gate is selected as input), as presented in Fig.~\ref{fig:strategies}~(d). $S_{4}$ allows two configurations as shown in Table~\ref{tab:locking:configurations}~(d).
Regardless of the key-bit value, all nodes remain connected; thereby forming a valid path.

Interestingly, driving each MUX in $S_{4}$ with an individual key is also a viable option. However, this configuration is susceptible to a simple reduction attack in case the input nodes are of the single-output type, thereby enabling SAAM. In this specific case, for $\{k_{i}, k_{j}\} = \{0, 1\}$ the node $f_{j}$ remains unconnected, whereas for $\{k_{i}, k_{j}\} = \{1, 0\}$ the node $f_{i}$ remains unconnected. Therefore, the adversary just has to guess whether $\{k_{i}, k_{j}\}$ is $\{0, 0\}$ or $\{1, 1\}$. This is equivalent to using $S_{4}$ with only one key. Note that $S_{4}$ is the only strategy which is always applicable.

\subsection{Cost Model}\label{costmodel}
The cost of a particular locking strategy can be expressed in terms of the number of gates inserted per key $K$ of length $l$, where $l=\abs{K}$. The cost summary for all strategies is presented in Table~\ref{tab:cost-summary}, assuming the following:
\begin{itemize}
	\item A single key bit input $k_{i}\in{K}$ is used only once.
	\item Every MUX is implemented with the same number of gates denoted as $\abs{MUX}$. Typically, a multiplexer is implemented by using one inverter, two AND gates, and one OR gate, i.e., $\abs{MUX} = 4$.
\end{itemize}
This model can be used to easily steer the final cost in the D-MUX locking procedure depending on the available strategies. 
%

\begin{table}[t]\scriptsize
	\centering
	\caption{Cost summary for D-MUX locking strategies.}
	\label{tab-size}
	\begin{tabular}{c|c|c}
		\hline
		\textbf{Locking Strategy}& \textbf{Number of Gates} & \textbf{Min. Key Bits per Iteration} \\ \hline
		\rowcolor{gray!20}\textbf{$S_{1}$} & $l\cdot{\abs{MUX}}$ & 2\\\hline
		\textbf{$S_{2}, S_{3}$} & $l\cdot{\abs{MUX}}$ & 1\\\hline
		\rowcolor{gray!20}\textbf{$S_{4}$} & $l\cdot{(2\cdot{\abs{MUX}})}$ & 1\\\hline
	\end{tabular}
	\label{tab:cost-summary}
\end{table}
\subsection{The D-MUX Locking Scheme}
In this section, we discuss the construction of the D-MUX LL scheme. The scheme is presented in Algorithm~\ref{alg:D-MUX}. The scheme takes the following inputs: a set of available strategies $L_{s}$, the correct key vector $K$, the original netlist $Net$, the maximum input node iteration variable $I_{max}$, and the maximum output node iteration variable $O_{max}$. Hereby, $I_{max}$ defines the maximum number of reselections of input node pairs and $O_{max}$ the maximum number of reselections of output node pairs for already selected input nodes. Note that $S_{4}$ is always available to the locking algorithm, therefore it is not included in $L_{s}$. The final output is the locked netlist. 

\begin{algorithm}[t]\scriptsize
	\caption{D-MUX locking scheme.}
	\label{alg:D-MUX}
	\begin{algorithmic}[1]
		\Require{Available strategies $L_{s}$, key $K$, netlist $Net$, max input node iterations $I_{max}$, max output node iterations $O_{max}$}
		\Ensure{Locked netlist}
		\State $F_{single} \gets$ ExtractSingleOutputNodes($Net$)
		\State $F_{multi} \gets$ ExtractMultiOutputNodes($Net$)
		\State $K_{list} \gets$ ToList($K$)\Comment{Convert key to list}
		\While{$\abs{K_{list}}$ \textgreater{~0}}
		\LineComment{Select a candidate strategy}
		\State RandomShuffle($L_{s}$)
		\State $fallback \gets$ TRUE; $S_{sel} \gets \emptyset$
		\For{$S_{i}$ in $L_{s}$}
		\LineComment{Enough keys available?}
		\If{$S_{i}$ == $S_{1}~$\&\&$~$($\abs{K_{list}}$ \textless ~2)}
		\State \textbf{continue}
		\EndIf
		\LineComment{Enough nodes available?}
		\If{$S_{i}\in{\{S_{1},S_{2},S_{3}\}}~$\&\&$~$($\abs{F_{multi}}$ \textless ~2)}
		\State \textbf{continue}
		\EndIf
		\State $S_{sel} \gets S_{i};~fallback \gets$ FALSE
		\State \textbf{break}
		\EndFor
		\If{$fallback$}
		\State $S_{sel} \gets{S_{4}}$
		\EndIf
		\LineComment{Search for valid input/output nodes}
		\State $\{\{f_{i}, f_{j}\}, \{g^{i}, g^{j}\},done\} \gets$
		\myindent{2} FindPairs($S_{sel}, F_{single}, F_{multi},I_{max},O_{max}$)
		\If{$!done$} \Comment{Check if the search was successful}
		\State \textbf{continue}
		\EndIf
		\LineComment{Apply selected strategy}
		\State $K_{i,j} \gets$ GetAndRemoveFrom($K_{list}, S_{sel}$)
		\State $\{M_{i,j}\}\gets $CoupleToMUXs($f_{i}, f_{j},g^{i}, g^{j}, K_{i,j}$)
		\State RegisterToNetlist($Net, \{M_{i,j}\}$)
		\EndWhile
		\State \textbf{return} $Net$
	\end{algorithmic}
\end{algorithm}

Before the main locking loop, the scheme prepares two separate sets: all single-output ($F_{single}$) and multi-output ($F_{multi}$) nodes (line 1 and 2). Furthermore, the key $K$ is represented as a list $K_{list}$ (line 3) to easily track the amount of used keys. The main locking loop repeats until no keys are left (line 4). Each iteration starts by randomly shuffling the available strategies (line 6). The shuffling ensures a random pick in the next step; all strategies in $L_{s}$ are iterated and multiple checks are performed to ensure the existence of enough key bits as well input nodes of the desired type (line 8 to 19).

Once a candidate $S_{i}$ is selected, the scheme proceeds with finding a valid pair of nodes (line 24). If successful, based on the nature of $S_{i}$, the scheme performs the following steps: retrieve the necessary key bits (set $K_{i,j}$, line 29) and couple all nodes through one or multiple MUXs (set $\{M_{i,j}\}$, line 30). Finally, all changes are registered in the netlist $Net$ (line 31).  

\begin{algorithm}[t]\scriptsize
	\caption{FindPairs Function}
	\label{alg:find-pair}
	\begin{algorithmic}[1]
		\Require{Strategy $S_{i}$, Set of single-output nodes $F_{single}$, set of multi-output nodes $F_{multi}$, max input node iterations $I_{max}$, max output node iterations $O_{max}$}
		\Ensure{Valid input nodes $\{f_{j}, f_{i}\}$, valid output nodes $\{g^{j}, g^{i}\}$, success indicator $done$}
		\State $\{F_{1},F_{2}\} \gets \{\emptyset\}$
		\LineComment{Select correct nodes}
		\If{$S_{i}\in{\{S_{1},S_{2}\}}$}
		\State $\{F_{1}, F_{2}\} \gets{\{F_{multi},F_{multi}\}}$
		\ElsIf{$S_{i}\in{\{S_{3}\}}$}
		\State $\{F_{1}, F_{2}\} \gets{\{F_{multi},F_{single}\}}$
		\Else
		\State $\{F_{1}, F_{2}\} \gets{\{F_{single}\cup{F_{multi}},F_{single}\cup{F_{multi}}\}}$
		\EndIf
		\LineComment{Prepare input and output node references}
		\State $\{f_{i}, f_{j}, g^{i}, g^{j}\}\gets{\emptyset};~done \gets$ FALSE
		\For{$iter_{in} = 0$ to $I_{max}$}
		\LineComment{Select first and second input node}
		\State $\{f_{i},f_{j}\} \gets$ $\{$RndSel($F_{1})$, RndSel($F_{2}$)$\}$)
		\LineComment{Select output nodes}
		\For{$iter_{out} = 0$ to $O_{max}$}
		\State $\{g^{i},g^{j}\} \gets \{$RndSel(OutsOf($f_{i}$)), RndSel(OutputsOf($f_{j}$))$\}$
		\State $\{R_{1},R_{2}\} \gets\{$IsInOutCone($f_{j}, g^{i}$), IsInOutCone($f_{i}, g^{j}$)$\}$
		\If{$(g^{i}~!=~g^{j})~$\&\&$~!R_{1}~$\&\&$~!R_{2}$}
		\State $done\gets$ TRUE
		\State \textbf{break}
		\EndIf
		\EndFor
		\If{$done$}
		\State\textbf{break}
		\EndIf		
		\EndFor
		\State \textbf{return} $\{\{f_{i}, f_{j}\}, \{g^{i}, g^{j}\},done\}$
	\end{algorithmic}
\end{algorithm}
\textbf{Scheme Variants:}~$L_{s}$ enables the creation of a generalized (gD-MUX) and enhanced (eD-MUX) scheme variant. In eD-MUX, $L_{s}=\{S_{1}, S_{2}, S_{3}\}$. Thus, the scheme first tries to use the less costly strategies if possible. In case none of these strategies are viable, the scheme \textit{falls back} to $S_{4}$ (line 21). In gD-MUX, $L_{s}=\emptyset$. This forces the scheme to exclusively use $S_{4}$. The difference between gD-MUX and eD-MUX mirrors the cost discrepancy between the cases when the target netlist only supports $S_{4}$ (worst case) and when it supports all other strategies (best case).


\textbf{Node Selection:} In all strategies, the input and output nodes must be carefully selected to avoid cycles. Therefore, D-MUX selects the nodes (line 24) according to the function \emph{FindPairs}($S_{i},F_{single}, F_{multi}, I_{max}, O_{max}$) described in Algorithm~\ref{alg:find-pair}. The function takes five inputs: the locking strategy $S_{i}$, the set of single-output nodes $F_{single}$, the set of multi-output nodes $F_{multi}$, and the maximum iteration variables $I_{max}$ and $O_{max}$. The output consists of the valid input ($\{f_{i}, f_{j}\}$) and output nodes ($\{g^{i}, g^{j}\}$) as well as an indicator ($done$) if the search has been successful. The function works as follows. Two node sets are assigned ($F_{1}$ and $F_{2}$) depending on the requirements of $S_{i}$ (line 1 to 9). These sets are used for the input node selection in the rest of the function. The selection loop (line 12) is repeated until two valid input and output nodes are found or the maximum iteration is reached. In each iteration, two candidate input nodes ($f_{i}$ and $f_{j}$) are randomly selected from $F_{1}$ and $F_{2}$, respectively. Next, the function tries to find two valid output nodes. The validity is defined by three requirements (line 19): $(i)$ the nodes are different, $(ii)$ $f_{j}$ is not in the output cone of $g^{i}$, and ($iii$) $f_{i}$ is not in the output cone of $g^{j}$. $(ii)$ and $(iii)$ prevent the creation of cycles. The node selection is limited to a given amount of iterations, since a valid selection is not always possible.
If a selection is not made, the function returns an indication that the search has failed. In this case, the algorithm repeats the search.

The selection returns two output nodes; even though some strategies require only one node. However, since every node has at least one output, this does not impact the final result. 

\section{Resilience Evaluation}\label{resilience:eval}
This section evaluates D-MUX against all relevant attacks, as discussed in Section~\ref{attacksoverview}. For the evaluation, we used the ISCAS'85~\cite{iscas2}, ITC'99~\cite{ITC99}, and RISC-V Ariane core~\cite{ariane2019} benchmarks listed in Table~\ref{tab:benchmarks}. 
All evaluations have been performed on an AMD Ryzen 9 3900X processor with 64GB of RAM and an Nvidia GeForce RTX 2080 Ti graphic card.


\begin{table}[b]\scriptsize
	\caption{Benchmark circuits used for evaluation.}
	\centering
	\subfloat[ISCAS'85]{	\tabcolsep=0.11cm
		\begin{tabular}{cc}\hline
			\textbf{IC} & \textbf{\#Gates}  \\
			\hline
			\rowcolor{gray!20}	{c1355} & 546  \\
			{c1908} & 880 \\
			\rowcolor{gray!20}{c2670} & 1193 \\
			{c3540} & 1669  \\
			\rowcolor{gray!20}{c5315} & 2307  \\
			{c6288} & 2416 \\
			\rowcolor{gray!20}	{c7552} & 3512  \\ \hline
	\end{tabular}}\hfil\subfloat[ITC'99]{	\tabcolsep=0.11cm
		\begin{tabular}{ccc}\hline
			\textbf{IC} & \textbf{\#Gates} &\textbf{\#FFs} \\
			\hline
			\rowcolor{gray!20}{b15}  & 6931 & 447  \\
			{b21}  & 7931 & 494  \\
			\rowcolor{gray!20}{b22} & 12128 & 709 \\
			{b17} & 21191 & 1407  \\
			\rowcolor{gray!20}{b18}  & 49293 & 3308\\
			{b19} & 98726 & 6618 \\ \hline
	\end{tabular}}\hfil	\subfloat[Ariane RISC-V]{	\tabcolsep=0.11cm
		\begin{tabular}{cc}\hline
			\textbf{IC} & \textbf{\#Gates}  \\
			\hline
			\rowcolor{gray!20}	{iscan}  & 240  \\
			{commit}  & 1584  \\
			\rowcolor{gray!20}	{brunit} & 1655  \\
			{decoder} & 2169  \\
			\rowcolor{gray!20}	{pcselect} & 3333 \\
			{brpredict} & 4669  \\
			\rowcolor{gray!20}	{alu}  & 7412 \\\hline
	\end{tabular}}
	\label{tab:benchmarks}
\end{table}

\subsection{SWEEP}

\textbf{Attack Setup:}~To evaluate D-MUX using SWEEP, we applied the SWEEP tool provided by~\cite{sweepAttack2019}. The setup is as follows. First, all benchmarks from Table~\ref{tab:benchmarks}~(a) have been copied 100 times and locked with random 64-bit keys; generating a data set of 700 locked benchmarks. Second, for each benchmark, the data set has been divided into a test set consisting of the 100 locked targets and a training set consisting of the remaining 600 other benchmarks. The attack is repeated for a range of margins, i.e., $m\in [0.0, 0.01]$ with a step of size $6.25\cdot{10^{-3}}$. This is repeated for both D-MUX variants.
\begin{figure}[t]
	\vspace{-0.2in}
	\centering
	\subfloat[gD-MUX]{
		\begin{tikzpicture}[scale=0.5]
			\pgfplotsset{
				compat=1.11,
				legend image code/.code={
					\draw[mark repeat=2,mark phase=2]
					plot coordinates {
						(0cm,0cm)
						(0.5cm,0cm)        
						(1cm,0cm)       
					};%
				}
			},
			\begin{axis}[
				enlarge x limits={abs=0.5cm},
				legend style={at={(0.7,0.83)},anchor=north, legend style={font=\large}, legend cell align={left}},
				legend columns=1,
				ymajorgrids,
				yminorgrids,
				ymin=-10,ymax=110,
				extra y ticks={25, 75},
				ylabel style={at={(-0.18,0.5)},anchor=north},        
				ylabel={Percentage (\%)},
				xmin=0,
				xmax=1,
				ylabel style ={font=\large},
				xlabel style ={font=\large},
				xlabel={Margin (\%)},
				major x tick style = {black, thick},
				major y tick style = {black, thick},
				minor tick length=2ex,
				width=8.5cm, height=4cm, 
				tick label style={font=\large} 
				]	
				%
				
				\addplot[dashdotted, mark size=1.5pt,every mark/.append style={solid}, mark=square*,mark options={solid, scale=1.5}, black, line width=1pt] table[x index=0,y index=1,col sep=comma] {average_Prec.txt};	
				
				\addplot[mark size=1.5pt,every mark/.append style={solid}, mark=*, rwth2red!90, line width=1pt,mark options={scale=1.5}] table[x index=0,y index=1,col sep=comma] {average_AccGuessed.txt};	
				
				\addplot[densely dashed, mark=triangle*,mark options={solid, scale=1.5}, mark size=1.5pt,every mark/.append style={solid}, gray, line width=1pt] table[x index=0,y index=1,col sep=comma] {average_Acc.txt};	
				
				\legend{Precision, KPA, Accuracy}
			\end{axis}
		\end{tikzpicture}
	}
	\subfloat[eD-MUX]{
		\begin{tikzpicture}[scale=0.5]
			\pgfplotsset{
				compat=1.11,
				legend image code/.code={
					\draw[mark repeat=2,mark phase=2]
					plot coordinates {
						(0cm,0cm)
						(0.5cm,0cm)        
						(1cm,0cm)       
					};%
				}
			},
			\begin{axis}[
				enlarge x limits={abs=0.5cm},
				legend style={at={(0.7,0.83)},anchor=north, legend style={font=\large}, legend cell align={left}},
				legend columns=1,
				ymajorgrids,
				yminorgrids,
				ymin=-10,ymax=110,
				extra y ticks={25,75},
				ylabel style={at={(-0.18,0.5)},anchor=north},        
				ylabel={Percentage (\%)},
				xmin=0,
				xmax=1,
				ylabel style ={font=\large},
				xlabel style ={font=\large},
				xlabel={Margin (\%)},
				major x tick style = {black, thick},
				major y tick style = {black, thick},
				minor tick length=2ex,
				width=8.5cm, height=4cm, 
				tick label style={font=\large} 
				]			
				\addplot[dashdotted, mark size=1.5pt,every mark/.append style={solid}, mark=square*,mark options={solid, scale=1.5}, black, line width=1pt] table[x index=0,y index=1,col sep=comma] {average_Prec_opt.txt};
				
				\addplot[mark size=1.5pt,every mark/.append style={solid}, mark=*, rwth1blue!70, line width=1pt,mark options={scale=1.5}] table[x index=0,y index=1,col sep=comma] {average_AccGuessed_opt.txt};	
				
				\addplot[densely dashed, mark=triangle*,mark options={solid, scale=1.5}, mark size=1.5pt,every mark/.append style={solid}, gray, line width=1pt] table[x index=0,y index=1,col sep=comma] {average_Acc_opt.txt};

				\legend{Precision, KPA, Accuracy}
			\end{axis}
		\end{tikzpicture}
	}
	\caption{SWEEP attack evaluation on D-MUX.} 
	\label{fig:sweep-dmux}
\end{figure}
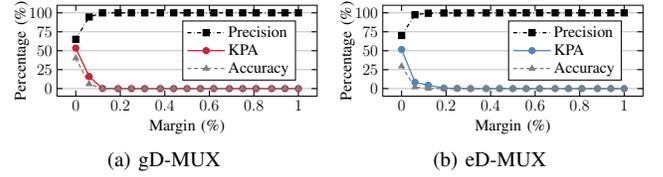
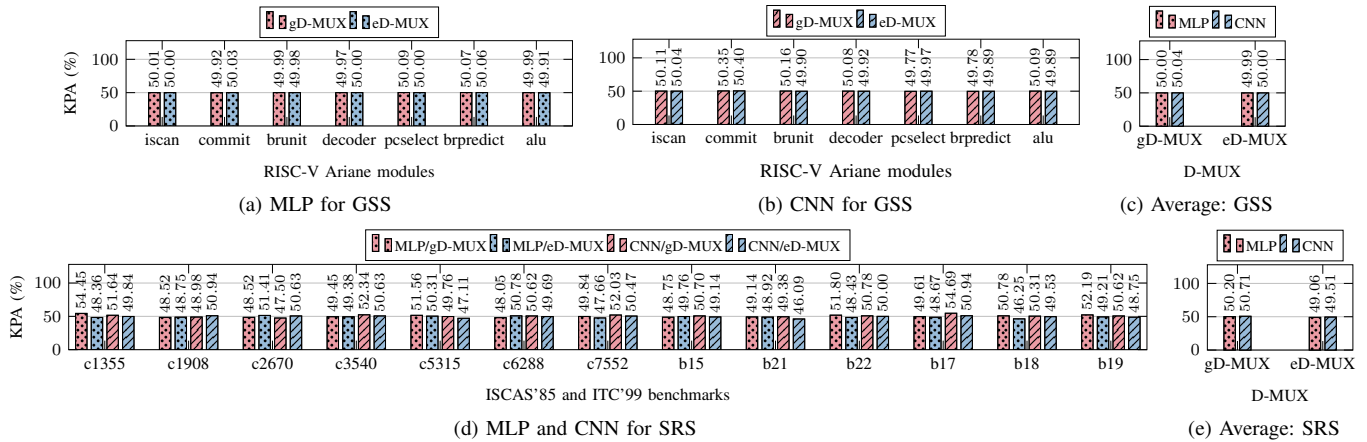
\begin{figure*}[ht]
	\centering

	\subfloat[MLP for GSS]{
		\begin{tikzpicture}[scale=0.8]
			\pgfplotstableread[row sep=\\,col sep=&]{			
				interval & eD-MUX & gD-MUX \\
				decoder & 50.00  & 49.97  \\
				pcselect & 50.00 & 50.09  \\ 
				brpredict  & 50.06 & 50.07  \\
				alu  & 49.91 & 49.99  \\
				iscan & 50.00 & 50.01 \\
				commit & 50.03 & 49.92 \\
				brunit & 49.98 & 49.99 \\
			}\mydata
			\begin{axis}[
				ylabel style={at={(0.011,0.5)},anchor=north}, 
				ybar,	legend columns=3,
				ymajorgrids = true,
				bar width=.19cm,
				width=9cm, height=3cm, 
				legend style={at={(0.5,1.4)},
					anchor=north,legend columns=-1,nodes={scale=0.7}},
				symbolic x coords={iscan, commit, brunit, decoder, pcselect, brpredict, alu},
				enlarge x limits={abs=0.6cm},
				ylabel style ={font=\footnotesize},
				xlabel style ={font=\footnotesize},
				every node near coord/.append style={
					anchor=north,
					yshift=2.5ex,
					xshift=-1.3ex,
					font=\scriptsize,
					rotate=90
				},
				every node near coord/.append style={
					/pgf/number format/fixed, 
					/pgf/number format/fixed zerofill,
					/pgf/number format/precision=2
				},
				yticklabel style = {font=\footnotesize},
				xticklabel style = {font=\footnotesize},
				ytick={0, 50, 100},	
				xtick align=inside,
				xtick=data,
				ymin=0,ymax=128,
				ylabel={KPA (\%)},
				xlabel={RISC-V Ariane modules},
				major x tick style = {black, thin},
				major y tick style = {black, thin},
				minor tick length=1ex,
				]
				\addplot[draw=black,fill=rwth2red!40, nodes near coords, postaction={pattern=crosshatch dots}] table[x=interval,y=gD-MUX]{\mydata};
				\addplot[draw=black,fill=rwth1blue!40, nodes near coords,postaction={pattern=crosshatch dots}] table[x=interval,y=eD-MUX]{\mydata};
				
				\legend{gD-MUX, eD-MUX}
			\end{axis}
		\end{tikzpicture}
	}\subfloat[CNN for GSS]{
		\begin{tikzpicture}[scale=0.8]
			\pgfplotstableread[row sep=\\,col sep=&]{			
				interval & eD-MUX & gD-MUX \\
				decoder & 49.92  & 50.08 \\
				pcselect & 49.97 & 49.77  \\ 
				brpredict  & 49.89 & 49.78 \\
				alu  & 49.89 &  50.09 \\
				iscan & 50.04 &  50.11\\
				commit & 50.40 & 50.35 \\
				brunit & 49.90 & 50.16\\
			}\mydata
			\begin{axis}[
				ybar,	legend columns=3,
				ymajorgrids = true,
				bar width=.19cm,
				width=9cm, height=3cm, 
				legend style={at={(0.5,1.4)},
					anchor=north,legend columns=-1,nodes={scale=0.7}},
				symbolic x coords={iscan, commit, brunit, decoder, pcselect, brpredict, alu},
				enlarge x limits={abs=0.6cm},
				ylabel style ={font=\small},
				xlabel style ={font=\small},
				every node near coord/.append style={
					anchor=north,
					yshift=2.5ex,
					xshift=-1.3ex,
					font=\scriptsize,
					rotate=90
				},
				yticklabel style = {font=\footnotesize},
				xticklabel style = {font=\footnotesize},
				every node near coord/.append style={
					/pgf/number format/fixed, 
					/pgf/number format/fixed zerofill,
					/pgf/number format/precision=2
				},
				ytick={0, 50, 100},	
				xtick align=inside,
				xtick=data,
				ymin=0,ymax=128,
				xlabel={RISC-V Ariane modules},
				major x tick style = {black, thin},
				major y tick style = {black, thin},
				minor tick length=1ex,
				]
				
				\addplot[draw=black,fill=rwth2red!40, nodes near coords, postaction={pattern=north east lines}] table[x=interval,y=gD-MUX]{\mydata};
				\addplot[draw=black,fill=rwth1blue!40, nodes near coords,postaction={pattern=north east lines}] table[x=interval,y=eD-MUX]{\mydata};
				\legend{gD-MUX, eD-MUX}
			\end{axis}
		\end{tikzpicture}
	} \subfloat[Average: GSS]{
		\begin{tikzpicture}[scale=0.8]
			\pgfplotstableread[row sep=\\,col sep=&]{			
				interval & MLP & CNN\\
				gD-MUX  & 50.00  & 50.04\\
				eD-MUX & 49.99 & 50.00\\   
			}\mydata
			\begin{axis}[
				ylabel style={at={(0.0075,0.5)},anchor=north}, 
				legend image post style={scale=1},
				ybar, legend columns=4,
				ymajorgrids = true,
				bar width=.19cm,
				width=4cm, height=3cm, 
				legend style={at={(0.5,1.41)},
					anchor=north,legend columns=-1,nodes={scale=0.7}},
				symbolic x coords={gD-MUX, eD-MUX},
				enlarge x limits={abs=0.5cm},
				ylabel style ={font=\footnotesize},
				xlabel style ={font=\footnotesize},
				every node near coord/.append style={
					anchor=north,
					yshift=2.5ex,
					xshift=-1.3ex,
					font=\scriptsize,
					rotate=90
				},
				yticklabel style = {font=\footnotesize},
				xticklabel style = {font=\footnotesize},
				every node near coord/.append style={
					/pgf/number format/fixed, 
					/pgf/number format/fixed zerofill,
					/pgf/number format/precision=2
				},
				ytick={0, 50, 100},	
				xtick align=inside,
				xtick=data,
				ymin=0,ymax=128,
				xlabel={D-MUX},
				major x tick style = {black, thin},
				major y tick style = {black, thin},
				minor tick length=1ex,
				]

				\addplot[draw=black,fill=rwth2red!40, nodes near coords,postaction={pattern=crosshatch dots}] table[x=interval,y=MLP]{\mydata};
				\addplot[draw=black,fill=rwth1blue!40, nodes near coords,postaction={pattern=north east lines}] table[x=interval,y=CNN]{\mydata};
				\legend{MLP, CNN}
			\end{axis}
		\end{tikzpicture}
	}	
	\vspace{-0.08in}
	\centering
	\subfloat[MLP and CNN for SRS]{
		\begin{tikzpicture}[scale=0.8]
			\pgfplotstableread[row sep=\\,col sep=&]{			
				interval & MLP/eD-MUX & MLP/gD-MUX & CNN/eD-MUX & CNN/gD-MUX \\
				c1355 & 48.359375 & 54.453125 & 49.84375 & 51.640625 \\
				c1908 & 48.75 & 48.515625 & 50.9375 & 48.984375\\   
				c2670 & 51.40625 & 48.515625  & 50.625&47.5 \\
				c3540 & 49.375 & 49.453125 & 50.625 &52.34 \\
				c5315 & 50.3125 & 51.5625 & 47.109375 &49.76 \\
				c6288 & 50.78125 & 48.046875 &49.6875  & 50.62\\
				c7552 & 47.65625 & 49.84375 & 50.46875 & 52.03125\\		
				b15   & 49.76  & 48.75 & 49.140625 & 50.70 \\
				b21   & 48.92  & 49.140625 & 46.09 & 49.375 \\
				b22   & 48.43  & 51.796875 & 50.00 & 50.78\\
				b17   & 48.67  & 49.609375 & 50.9375& 54.6875\\
				b18   & 46.25 & 50.78125 & 49.53125  & 50.3125\\
				b19   & 49.21  & 52.1875 & 48.75& 50.62\\
			}\mydata
			\begin{axis}[
				ylabel style={at={(0.0075,0.5)},anchor=north}, 
				legend image post style={scale=1},
				ybar, legend columns=4,
				ymajorgrids = true,
				bar width=.19cm,
				width=19.5cm, height=3cm, 
				legend style={at={(0.5,1.41)},
					anchor=north,legend columns=-1,nodes={scale=0.7}},
				symbolic x coords={c1355, c1908, c2670, c3540, c5315, c6288, c7552, b15, b21, b22, b17, b18, b19},
				enlarge x limits={abs=0.6cm},
				ylabel style ={font=\footnotesize},
				xlabel style ={font=\footnotesize},
				every node near coord/.append style={
					anchor=north,
					yshift=2.5ex,
					xshift=-1.3ex,
					font=\scriptsize,
					rotate=90
				},
				yticklabel style = {font=\footnotesize},
				xticklabel style = {font=\footnotesize},
				every node near coord/.append style={
					/pgf/number format/fixed, 
					/pgf/number format/fixed zerofill,
					/pgf/number format/precision=2
				},
				ytick={0, 50, 100},	
				xtick align=inside,
				xtick=data,
				ymin=0,ymax=128,
				ylabel={KPA (\%)},
				xlabel={ISCAS'85 and ITC'99 benchmarks},
				major x tick style = {black, thin},
				major y tick style = {black, thin},
				minor tick length=1ex,
				]

				\addplot[draw=black,fill=rwth2red!40, nodes near coords,postaction={pattern=crosshatch dots}] table[x=interval,y=MLP/gD-MUX]{\mydata};
				\addplot[draw=black,fill=rwth1blue!40, nodes near coords,postaction={pattern=crosshatch dots}] table[x=interval,y=MLP/eD-MUX]{\mydata};
				\addplot[draw=black,fill=rwth2red!40, nodes near coords, postaction={pattern=north east lines}] table[x=interval,y=CNN/gD-MUX]{\mydata};
				\addplot[draw=black,fill=rwth1blue!40, nodes near coords,postaction={pattern=north east lines}] table[x=interval,y=CNN/eD-MUX]{\mydata};
				\legend{MLP/gD-MUX, MLP/eD-MUX,CNN/gD-MUX, CNN/eD-MUX}
				
			\end{axis}
		\end{tikzpicture}
	}\subfloat[Average: SRS]{
		\begin{tikzpicture}[scale=0.8]
			\pgfplotstableread[row sep=\\,col sep=&]{			
				interval & MLP & CNN\\
				gD-MUX  & 50.20  & 50.71\\
				eD-MUX & 49.06 & 49.51\\   
			}\mydata
			\begin{axis}[
				ylabel style={at={(0.0075,0.5)},anchor=north}, 
				legend image post style={scale=1},
				ybar, legend columns=4,
				ymajorgrids = true,
				bar width=.19cm,
				width=4cm, height=3cm, 
				legend style={at={(0.5,1.41)},
					anchor=north,legend columns=-1,nodes={scale=0.7}},
				symbolic x coords={gD-MUX, eD-MUX},
				enlarge x limits={abs=0.5cm},
				ylabel style ={font=\footnotesize},
				xlabel style ={font=\footnotesize},
				every node near coord/.append style={
					anchor=north,
					yshift=2.5ex,
					xshift=-1.3ex,
					font=\scriptsize,
					rotate=90
				},
				yticklabel style = {font=\footnotesize},
				xticklabel style = {font=\footnotesize},
				every node near coord/.append style={
					/pgf/number format/fixed, 
					/pgf/number format/fixed zerofill,
					/pgf/number format/precision=2
				},
				ytick={0, 50, 100},	
				xtick align=inside,
				xtick=data,
				ymin=0,ymax=128,
				xlabel={D-MUX},
				major x tick style = {black, thin},
				major y tick style = {black, thin},
				minor tick length=1ex,
				]

				\addplot[draw=black,fill=rwth2red!40, nodes near coords,postaction={pattern=crosshatch dots}] table[x=interval,y=MLP]{\mydata};
				\addplot[draw=black,fill=rwth1blue!40, nodes near coords,postaction={pattern=north east lines}] table[x=interval,y=CNN]{\mydata};
				\legend{MLP, CNN}
			\end{axis}
		\end{tikzpicture}
	}	
	\caption{SnapShot attack evaluation on D-MUX for the generalized set and self-referencing scenario.}
	\label{fig:evalsnapshot}
	\vspace{-0.1in}
\end{figure*}

\textbf{Metrics:}~The attack was evaluated through accuracy, precision, and KPA (see Section~\ref{sweepdef}). The accuracy in SWEEP is calculated using the whole key length, regardless of X values. The KPA only considers the bits that are different from X.

\textbf{Results:}~The average attack results across all benchmarks and margins are presented in Fig.~\ref{fig:sweep-dmux}. At $m=0$, the average KPA is approx. 50\% as expected. However, the accuracy is lower than 50\% as its formulation takes X values into account. As soon as the margin value is slightly increased ($m>0$), both the accuracy and the KPA fall to 0\% while the precision rises to 100\%. \textit{This is due to the fact that with an increased margin, SWEEP is not able to report any 'guessed' key bits (\textbf{all bits} are marked as X).} 
Based on the evaluation, we can conclude that SWEEP is not able to extract a meaningful correlation from the data regardless of the acceptable margin.

\subsection{Learning Resilience}

\subsubsection{ANT and RNT}
In ANT, \textbf{U} is able to identify the inserted MUX gates by following each key input. Thereby, the only observation that can be made is that both inputs to the MUX are AND gates. However, due to the nature of the locking strategies deployed by D-MUX, it is not possible to distinguish the true and the false gate driving the MUX. Even after many iterations of the game, the guessing accuracy will remain 50\%. Therefore, \textit{D-MUX passes ANT}, thereby being evaluated as \textit{possibly} learning resilient. The same conclusion is drawn in RNT as well. The reason why D-MUX passes both tests lies within the way the key value is mapped into the circuits; instead of adding key-bounded functionality, D-MUX reconfigures the existing one.

\subsubsection{SnapShot}
So far, SnapShot represents the only oracle-less ML-based attack on LL that is applicable to MUX-based locking. SnapShot predicts key bits based on labeled LVs that structurally represent a subcircuit associated to a key-bit value. To adapt SnapShot to D-MUX, we have to introduce a new LV extraction procedure that includes all structurally relevant components of the netlist for each key input.

\textbf{D-MUX Locality Vector:}~The LV must include all relevant data, which could potentially leak information about the key. For D-MUX, we can observe two cases: one or two MUX gates are inserted per key bit. Both cases decompose a netlist into multiple components as shown in Fig.~\ref{fig:snapshot-dmux-representation}: 
\begin{figure}[b]
	\centering
	\subfloat[$S_{1}$ to $S_{3}$]{
		\includegraphics[width=0.38\columnwidth]{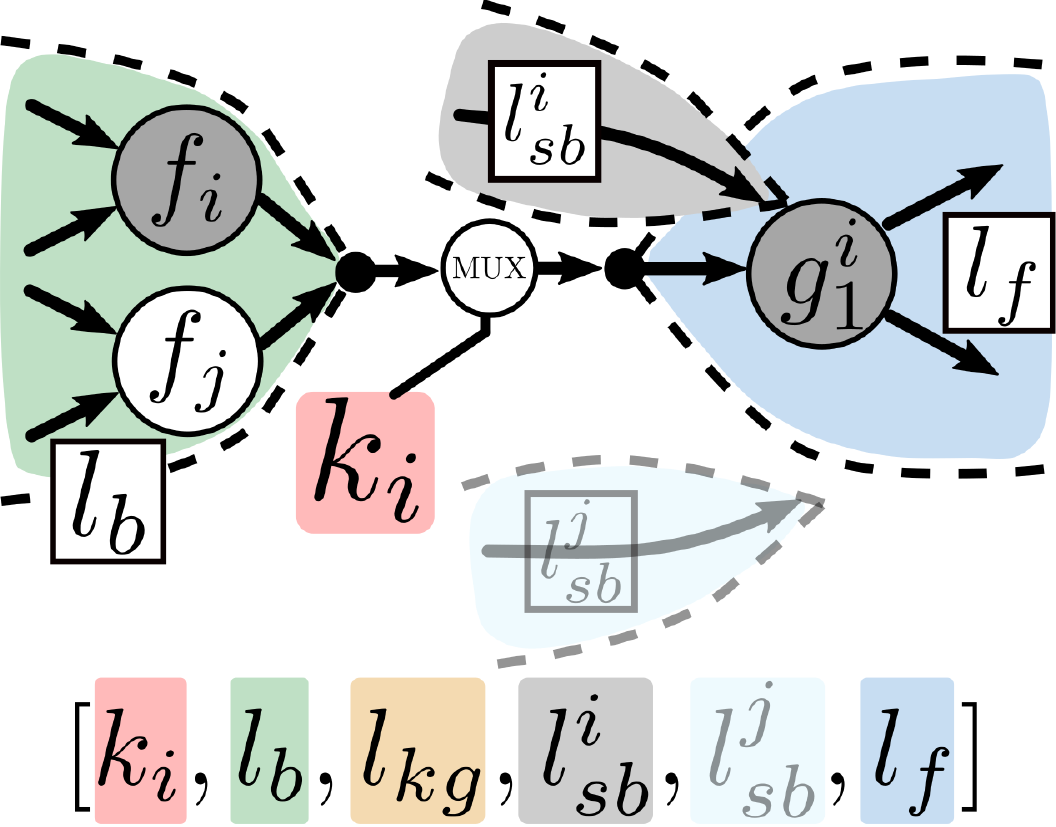}
	}\hfil
	\subfloat[$S_{4}$]{
		\includegraphics[width=0.38\columnwidth]{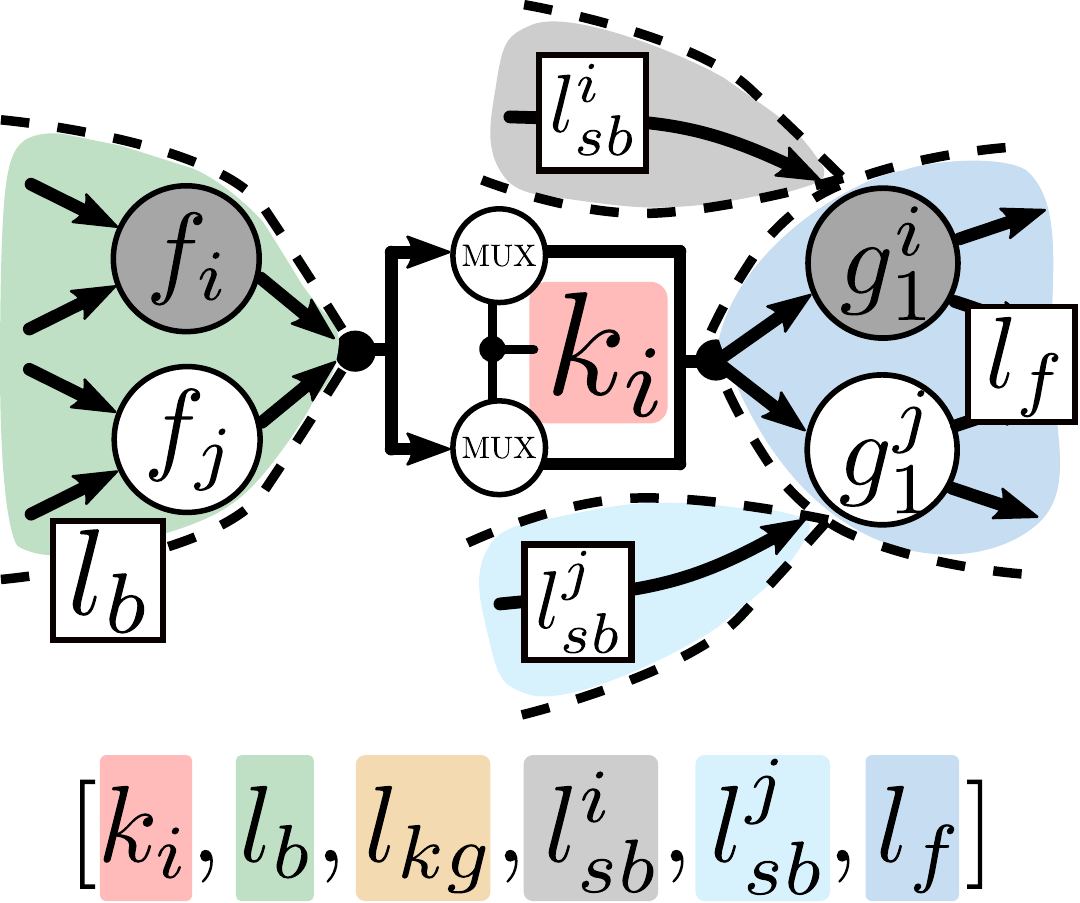}
	}
	\caption{Locality vectors for D-MUX.}
	\label{fig:snapshot-dmux-representation}
\end{figure}

\begin{itemize}
	\item $k_{i}$: the key-bit value (in case the vector is labeled).
	\item $l_{b}$: the backward path; the input cones of $f_{i}$ and $f_{j}$.
	\item $l_{kg}$: the key gate in the form of one MUX gate (regardless of whether one or two MUX gates are inserted).
	\item $l^{i}_{sb}$: the backward path specific to $g^{i}_{1}$.
	\item $l^{j}_{sb}$: the backward path specific to $g^{j}_{1}$ (if not existent, $l^{j}_{sb}$ is filled with zeros).
	\item $l_{f}$: the forward path; the output cones of $g_{i}$ and $g_{j}$.
\end{itemize} 
\begin{figure}[b]
	\centering
	\includegraphics[width=\columnwidth]{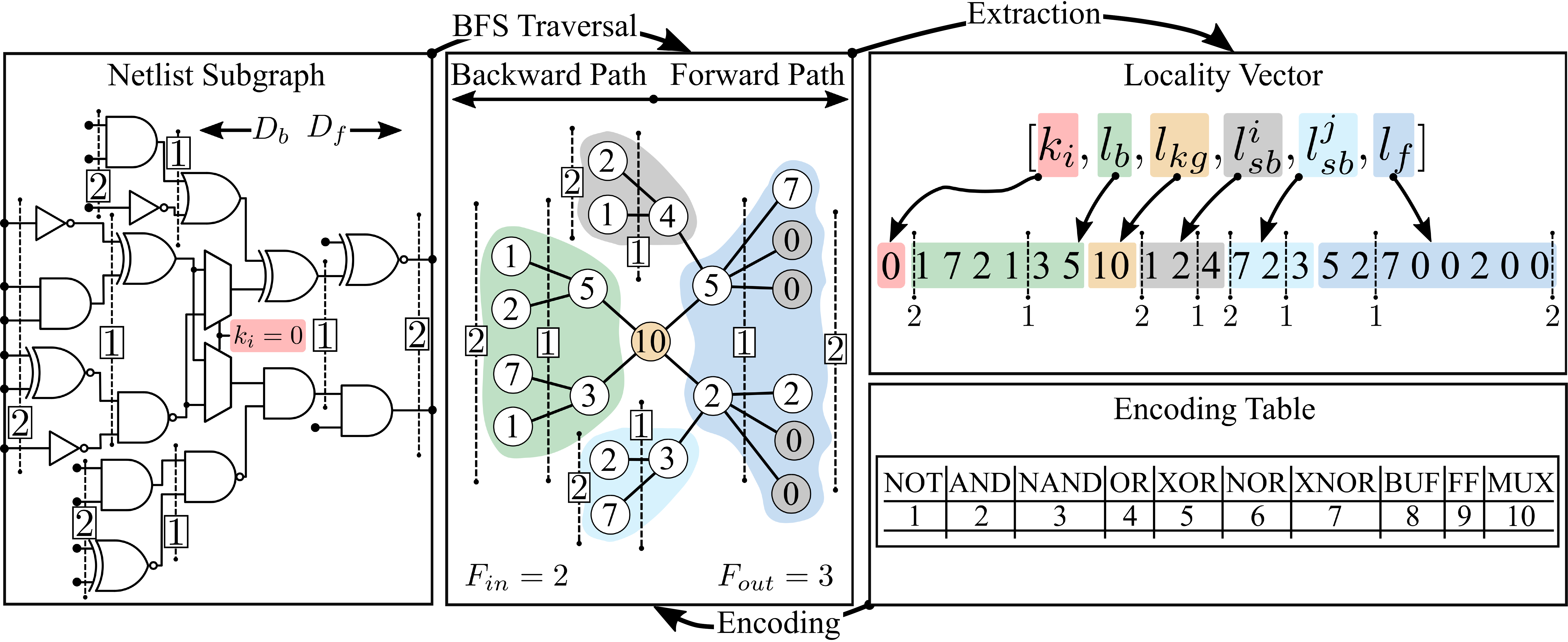}
	\caption{Example: locality vector extraction for D-MUX.}
	\label{fig:localityextractiondmux}
\end{figure}

\begin{figure*}[t]
	\centering
	\subfloat[c1355]{
			\begin{tikzpicture}[scale=0.5]
			\pgfplotsset{
				compat=1.11,
				legend image code/.code={
					\draw[mark repeat=2,mark phase=2]
					plot coordinates {
						(0cm,0cm)
						(0.5cm,0cm)        
						(1cm,0cm)       
					};%
				}
			},
			\begin{axis}[
				enlarge x limits={abs=0.4cm},
				legend style={at={(0.5,1.27)},anchor=north},
				legend columns=3,
				ymajorgrids,
				yminorgrids,
				xmajorgrids,
				ymin=0,ymax=4,
				ylabel style={at={(-0.135,0.5)},anchor=north},        
				ylabel={Post-synth. area [kGE]},
				xmin=0.4,
				xmax=3.0,
				ylabel style ={font=\large},
				xlabel style ={font=\large},
				xlabel={$T_{clk}$ [ns]},
				major x tick style = {black, thick},
				major y tick style = {black, thick},
				minor tick length=2ex,
				width=8.88cm, height=4.2cm, 
				tick label style={font=\large} 
				]

				\addplot[const plot,mark size=1.5pt,every mark/.append style={solid}, mark=*, black!80, line width=1pt,mark options={scale=1.5}] table[x index=0,y index=1,col sep=comma] {at_original_c1355.txt};	
				\label{p1}

				\addplot[const plot,mark size=1.5pt,every mark/.append style={solid}, mark=pentagon*, rwth1blue, line width=1pt,mark options={scale=1.5}] table[x index=0,y index=1,col sep=comma] {at_optimized_c1355.txt};
				\label{p2}
				\addplot[const plot,mark size=1.5pt,every mark/.append style={solid}, mark=triangle*, rwth2red, line width=1pt,mark options={scale=1.5}] table[x index=0,y index=1,col sep=comma] {at_generalized_c1355.txt};
				\label{p3}
				\addplot[only marks,mark=*,black!80,every mark/.append style={solid},mark options={scale=2,fill=white},text mark as node=true,forget plot] coordinates { (0.5,0.44)};
				\label{p4}
				\addplot[only marks,rwth1blue,mark=pentagon*,every mark/.append style={solid},mark options={scale=2,fill=white},text mark as node=true,forget plot] coordinates { (1.1, 0.66)};
				\label{p5}
				\addplot[only marks,rwth2red,mark=triangle*,every mark/.append style={solid},mark options={scale=2,fill=white},text mark as node=true, forget plot] coordinates { (1.2, 1.0 )};
				\label{p6}
				
				%
				%
				%
				
				
				\node [draw,fill=white] at (rel axis cs: 0.71, 0.63) {\shortstack[l]{
						\ref{p6} $AT^{opt}_{gD\text{-}MUX}~~(1.2, 1.0 )$ \\
						\ref{p5} $AT^{opt}_{eD\text{-}MUX}~(1.1, 0.66)$ \\
						\ref{p4} $AT^{opt}_{original}~~(0.50, 0.44)$}};
				
				\legend{orig., eD-MUX, gD-MUX}
			\end{axis}
		\end{tikzpicture}
	}
	\subfloat[Optimum AT]{
			\begin{tikzpicture}[scale=0.55]
			\pgfplotstableread[row sep=\\,col sep=&]{
				interval & eD-MUX & gD-MUX \\
				area & 47.26 & 73.38 \\
				power & 51.09 & 52.76 \\
				delay & 17.27 & 48.83 \\
			}\mydata
			\begin{axis}[
				ylabel style={at={(-0.008,0.5)},anchor=north}, 
				ybar,	legend columns=2,
				ymajorgrids = true,
				bar width=.5cm,
				width=7cm, height=4cm, 
				legend style={at={(0.5,1.27)},anchor=north},
				symbolic x coords={area, power, delay},
				enlarge x limits={abs=0.8cm},
				ylabel style ={font=\large},
				xlabel style ={font=\large},
				tick label style={font=\large}, 
				nodes near coords always on top/.style={
					scatter/position=absolute,
					positive value/.style={
						at={(axis cs:\pgfkeysvalueof{/data point/x},\pgfkeysvalueof{/data point/y})},
					},
					negative value/.style={
						at={(axis cs:\pgfkeysvalueof{/data point/x},0)},
					},
					every node near coord/.append style={
						font=\large,
						check values/.code={%
							\begingroup
							\pgfkeys{/pgf/fpu}%
							\pgfmathparse{\pgfplotspointmeta<0}%
							\global\let\result=\pgfmathresult
							\endgroup
							%
							%
							\pgfmathfloatcreate{1}{1.0}{0}%
							\let\ONE=\pgfmathresult
							\ifx\result\ONE
							\pgfkeysalso{/pgfplots/negative value}%
							\else
							\pgfkeysalso{/pgfplots/positive value}%
							\fi
						},
						check values,
						anchor=west,
						rotate=90,
					},
				},
				nodes near coords={
					\pgfmathprintnumber[fixed zerofill,precision=1]{\pgfplotspointmeta}
				},
				nodes near coords always on top,
				every node near coord/.append style={
					/pgf/number format/fixed, 
					/pgf/number format/fixed zerofill,
					/pgf/number format/precision=2
				},	
				xtick align=inside,
				xtick=data,
				ymin=0,ymax=160,
				ylabel={\% overhead},
				xlabel={APD},
				ytick={25, 50, 75, 100},
				major x tick style = {black, thin},
				major y tick style = {black, thin},
				minor tick length=1ex,
				]
				
				
				\addplot[draw=black,fill=rwth2red!40, nodes near coords,postaction={pattern=crosshatch dots}] table[x=interval,y=gD-MUX]{\mydata};
				\addplot[draw=black,fill=rwth1blue!40, nodes near coords,postaction={pattern=north east lines}] table[x=interval,y=eD-MUX]{\mydata};
				
				\legend{gD-MUX, eD-MUX}
			\end{axis}
		\end{tikzpicture}
	}	
	\subfloat[High Performance]{
			\begin{tikzpicture}[scale=0.55]
			\pgfplotstableread[row sep=\\,col sep=&]{
				interval & eD-MUX & gD-MUX \\
				area & 43.17 & 87.02 \\
				power & 28.82 & 45.39 \\
				delay & 17.66 & 36.79 \\
			}\mydata
			\begin{axis}[
				ylabel style={at={(-0.008,0.5)},anchor=north}, 
				ybar,	legend columns=2,
				ymajorgrids = true,
				bar width=.5cm,
				width=7cm, height=4cm, 
				legend style={at={(0.5,1.27)},anchor=north},
				symbolic x coords={area, power, delay},
				enlarge x limits={abs=0.8cm},
				ylabel style ={font=\large},
				xlabel style ={font=\large},
				tick label style={font=\large}, 
				nodes near coords always on top/.style={
					scatter/position=absolute,
					positive value/.style={
						at={(axis cs:\pgfkeysvalueof{/data point/x},\pgfkeysvalueof{/data point/y})},
					},
					negative value/.style={
						at={(axis cs:\pgfkeysvalueof{/data point/x},0)},
					},
					every node near coord/.append style={
						font=\large,
						check values/.code={%
							\begingroup
							\pgfkeys{/pgf/fpu}%
							\pgfmathparse{\pgfplotspointmeta<0}%
							\global\let\result=\pgfmathresult
							\endgroup
							%
							%
							\pgfmathfloatcreate{1}{1.0}{0}%
							\let\ONE=\pgfmathresult
							\ifx\result\ONE
							\pgfkeysalso{/pgfplots/negative value}%
							\else
							\pgfkeysalso{/pgfplots/positive value}%
							\fi
						},
						check values,
						anchor=west,
						rotate=90,
					},
				},
				nodes near coords={
					\pgfmathprintnumber[fixed zerofill,precision=1]{\pgfplotspointmeta}
				},
				nodes near coords always on top,
				every node near coord/.append style={
					/pgf/number format/fixed, 
					/pgf/number format/fixed zerofill,
					/pgf/number format/precision=2
				},	
				xtick align=inside,
				xtick=data,
				ymin=0,ymax=160,
				ylabel={\% overhead},
				xlabel={APD},
				ytick={25, 50, 75, 100},
				major x tick style = {black, thin},
				major y tick style = {black, thin},
				minor tick length=1ex,
				]
				
				\addplot[draw=black,fill=rwth2red!40, nodes near coords,postaction={pattern=crosshatch dots}] table[x=interval,y=gD-MUX]{\mydata};
				\addplot[draw=black,fill=rwth1blue!40, nodes near coords,postaction={pattern=north east lines}] table[x=interval,y=eD-MUX]{\mydata};
				
				\legend{gD-MUX, eD-MUX}
			\end{axis}
		\end{tikzpicture}
	}
	\subfloat[Low Performance]{
			\begin{tikzpicture}[scale=0.55]
			\pgfplotstableread[row sep=\\,col sep=&]{
				interval & eD-MUX & gD-MUX \\
				area & 22.73 & 51.34 \\
				power & 32.43 & 71.3 \\
			}\mydata
			
			\begin{axis}[
				ylabel style={at={(-0.008,0.5)},anchor=north}, 
				ybar,	legend columns=2,
				ymajorgrids = true,
				bar width=.5cm,
				width=5cm, height=4cm, 
				legend style={at={(0.5,1.27)},anchor=north},
				symbolic x coords={area, power},
				enlarge x limits={abs=0.8cm},
				ylabel style ={font=\large},
				xlabel style ={font=\large},
				tick label style={font=\large}, 
				nodes near coords always on top/.style={
					scatter/position=absolute,
					positive value/.style={
						at={(axis cs:\pgfkeysvalueof{/data point/x},\pgfkeysvalueof{/data point/y})},
					},
					negative value/.style={
						at={(axis cs:\pgfkeysvalueof{/data point/x},0)},
					},
					every node near coord/.append style={
						font=\large,
						check values/.code={%
							\begingroup
							\pgfkeys{/pgf/fpu}%
							\pgfmathparse{\pgfplotspointmeta<0}%
							\global\let\result=\pgfmathresult
							\endgroup
							%
							%
							\pgfmathfloatcreate{1}{1.0}{0}%
							\let\ONE=\pgfmathresult
							\ifx\result\ONE
							\pgfkeysalso{/pgfplots/negative value}%
							\else
							\pgfkeysalso{/pgfplots/positive value}%
							\fi
						},
						check values,
						anchor=west,
						rotate=90,
					},
				},
				nodes near coords={
					\pgfmathprintnumber[fixed zerofill,precision=1]{\pgfplotspointmeta}
				},
				nodes near coords always on top,
				every node near coord/.append style={
					/pgf/number format/fixed, 
					/pgf/number format/fixed zerofill,
					/pgf/number format/precision=2
				},	
				xtick align=inside,
				xtick=data,
				ymin=0,ymax=160,
				ylabel={\% overhead},
				xlabel={AP},
				major x tick style = {black, thin},
				major y tick style = {black, thin},
				ytick={25, 50, 75, 100},
				minor tick length=1ex,
				]
				
				\addplot[draw=black,fill=rwth2red!40, nodes near coords,postaction={pattern=crosshatch dots}] table[x=interval,y=gD-MUX]{\mydata};
				
				\addplot[draw=black,fill=rwth1blue!40, nodes near coords,postaction={pattern=north east lines}] table[x=interval,y=eD-MUX]{\mydata};
				
				\legend{gD-MUX, eD-MUX}
			\end{axis}
		\end{tikzpicture}
	}
	
	\caption{D-MUX cost evaluation.}
	\label{fig:apd}
	\vspace{-0.1in}
\end{figure*}
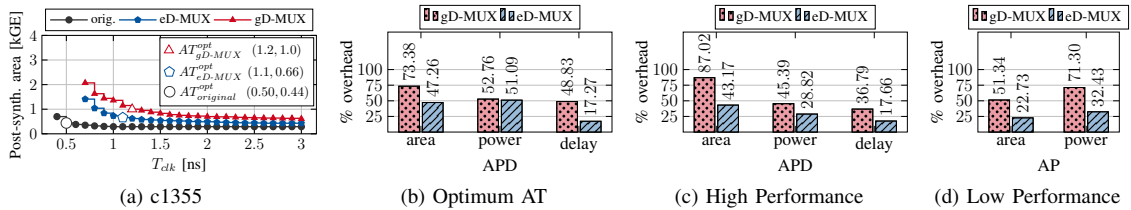
The extraction traverses each backward and forward path until a selected depth is reached, thereby applying the fan-in and fan-out values as suggested in~\cite{sisejkovic2020challenging}.
By applying the BFS, one locality vector can be extracted for each key input. One example is given in Fig.~\ref{fig:localityextractiondmux}. Here, the maximum forward and backward depth is set to 5 ($D_{b}=5$ and $D_{f}=5$). Each section of the vector represents one of the components listed above. 

\textbf{Attack Scenarios:} We evaluated both D-MUX variants using both attack scenarios as discussed in Section~\ref{snapshotintro}. In GSS, the training set is assembled using all benchmarks from Table~\ref{tab:benchmarks}~(a) and (b). Each benchmark is copied and locked 1,000 times using 64-bit keys, resulting in 13,000 locked netlists and 832,000 labeled localities. The test set consists of all benchmarks from Table~\ref{tab:benchmarks}~(c), each locked 1,000 times, yielding $448,000$ unlabeled localities. SRS generates a training set by relocking the each target 1,000 times to predict the key of the initial target. This is repeated 20 times to get an average value for all benchmarks in Table~\ref{tab:benchmarks}~(a) and (b).

\textbf{ML Model:}~We adapted the Mutli-layer Perceptron (MLP) and the CNN model from~\cite{sisejkovic2020challenging} to accommodate the higher complexity of D-MUX as follows; ($i$) in MLP, the input layer has a higher number of nodes (same as locality vector size), $(ii)$ we increased the number of available internal layers in the CNN evolution to 14, and $(iii)$ the number of epochs is set to 100 in the exploratory phase of the CNN.

\textbf{Results:}~All evaluation results are presented in Fig.~\ref{fig:evalsnapshot}. For all scenarios (MLP/CNN and GSS/SRS) the key prediction accuracy is consistently around 50\%. 
Thus, we can conclude that D-MUX is efficient in protecting against SnapShot, thereby forcing the attack to perform random guesses about the key. 

\textbf{Evolved CNNs:}~The evolved, best-performing networks tend to have few or no convolutional layers, thus degrading the CNN to a basic MLP. This suggests that no feature extraction is possible and a learning-by-heart approach yields best results.

\section{Cost Evaluation} \label{costeval}

The following evaluation gathers a realistic cost analysis using the UMC 90 nm CMOS process.
based on the ISCAS'85 and ITC'99 benchmarks from Table~\ref{tab:benchmarks}. The Synopsys Design Compiler (DC) was used for logic synthesis.



All benchmarks were locked with 64-bit keys for gD-MUX and eD-MUX separately. Each benchmark is synthesized for a range of clock periods ($T_{clk}$) to generate the Area-Timing (AT) graph. As an example, we only present the AT graph for the benchmark c1355 in Fig.~\ref{fig:apd}~(a). 
The graph includes the optimum AT point ($AT^{opt}$) for the original netlist and the locked variants. $AT^{opt}$ is the Pareto-optimal point with the most \textit{efficient} implementation (minimum $Area\cdot{T_{clk}}$). The lowest $T_{clk}$ is determined by the critical path, while the highest $T_{clk}$ is selected to exhibit a reasonable AT curve. To include locking and synthesis variations, the average value across 20 synthesis rounds for 20 locked netlist instances is taken.

The average cost across all benchmarks are presented in Fig.~\ref{fig:apd}~(b)-(d). The power overhead is approximated using DC, thus it correlates with the area overhead increase. We present the evaluations at three comparisons points as follows.

\textbf{Optimum AT:}~The average Area-Power-Delay (APD) overhead for optimum AT points is presented in Fig.~\ref{fig:apd}~(b). This evaluation shows the average difference between \textit{the most efficient implementations} per benchmark. Hereby, gD-MUX and eD-MUX induce an average area overhead of 73.38\% and 47.26\% respectively. In terms of delay, eD-MUX exhibits only 17.27\% overhead compared to the 48.83\% of gD-MUX. This is due to the fact that the AT optima of eD-MUX are typically closer in terms of $T_{clk}$ to the original AT optima. 

\textbf{High performance:}~This evaluation shows the cost for \textit{the lowest possible $T_{clk}$} (Fig.~\ref{fig:apd}~(c)). The results are similar to the optimum AT case. Most importantly, the delay overhead is limited to 36.79\% for gD-MUX and 17.66\% for eD-MUX. This suggests that in practice the locking mechanisms have a relatively low impact on the critical path of the designs.

\textbf{Low performance:}~This evaluation looks at the case \textit{when no excessive synthesis-optimizations have been applied yet} since enough margin in terms of $T_{clk}$ is still available, yielding a fair comparison (Fig.~\ref{fig:apd}~(d)). The longest shown clock period has been selected for comparison for each benchmarks (e.g., $T_{clk}=3$ ns in Fig.~\ref{fig:apd}~(a)). As before, gD-MUX exhibits a higher area cost (51.34\%) compared to eD-MUX (22.73\%). 
 

In conclusion, the evaluation shows that gD-MUX is costlier in terms of APD compared to eD-MUX for all benchmarks; hence offering a more cost-effective option.

\section{Analysis of Related Work}\label{related-work}
The following overview of related work focuses on four categories: MUX-based, routing-based, lookup table (LUT)-based, and ML-resilient LL. Hereby, we offer an analysis of potential structural leakage points in these schemes and indicate new perspectives on improving LL in the ML era.

 \begin{figure}[b]
 	\centering
 	\includegraphics[width=\columnwidth]{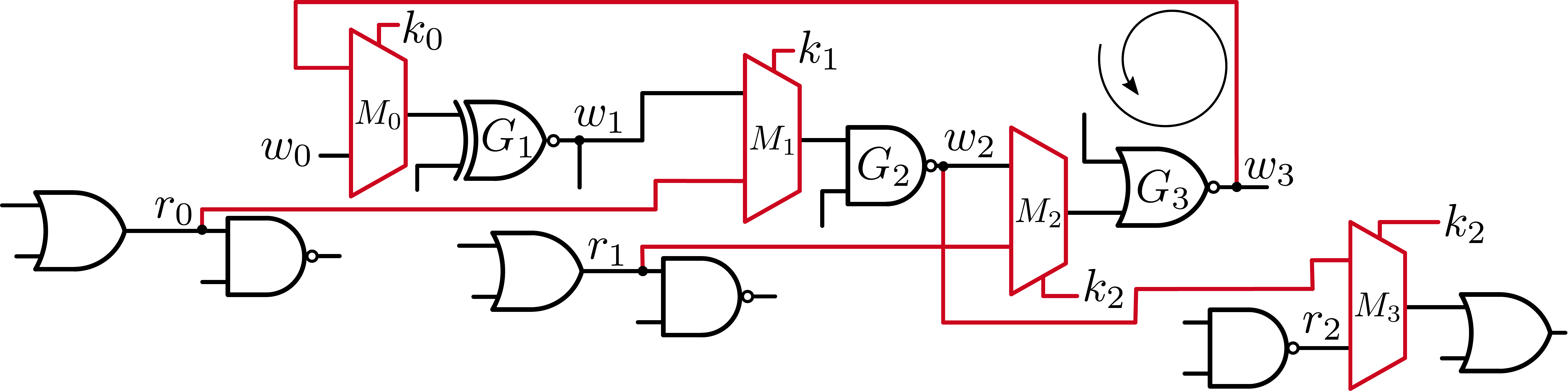}
 	\caption{Cyclic logic locking; false wires are marked red.}
 	\label{fig:cyclic_obf}
 \end{figure}
 
\textbf{MUX-Based LL:}~As mentioned in Section~\ref{saam:attack}, existing MUX-based LL techniques are susceptible to SAAM~\cite{plaza2015solving,logicConeAnalysis2015, fault2015}. However, a seemingly structurally secure scheme is known as cyclic obfuscation~\cite{cyclicObf2017}. This scheme has been designed to thwart SAT-based attacks by creating combinational cycles (later successfully challenged by CycSAT~\cite{CycSAT17}). To showcase its mode of operation, let us consider the locked example in Fig.~\ref{fig:cyclic_obf} (exact replica from~\cite{cyclicObf2017}). Here, all false wires are marked red. The scheme receives two inputs: the number of cycles and the cycle length. To insert a cycle, the scheme first searches for a path of a selected length between two gates; for example, the path $(w_{0}, w_{3})$. First, a feedback wire is added between these endpoints (MUX $M_{0}$). Second, all edges between $(w_{0}, w_{3})$ must be made \emph{removable} to disable the identification of the inserted cycle. This procedure is done by traversing all gates $G_{i}$ alongside the path, thereby two MUXs are added if the fan-out($G_{i}$) = 1 ($M_{2}$ and $M_{3}$) and, otherwise, only one MUX is inserted ($M_{1}$). Note that it is suggested that the same key input ($k_{2}$) drives both MUXs in the former case. With this configuration, all edges of the created cycle are removable. Even though this scheme checks for dangling wires, it still suffers from major leakage points. First, cyclic locking avoids dangling wires by checking the fan-out size of the gates driving the added MUXs. However, this check is not performed for the MUXs that are inserted \emph{outside} the path, i.e., for false wires. The presented example showcases this fault; the wire $r_{2}$ is only driving $M_{3}$, thus it has to be the true wire. Moreover, this error can also occur for the insertion of the MUX that creates the initial cycle. In the example, the fan-out of the gate that outputs $w_{0}$ is not checked. Evidently, if its fan-out is 1, $w_{0}$ must be the \emph{true} wire. These critical cases have not been taken into account in the proposal~\cite{cyclicObf2017}. Furthermore, we have verified this leakage in the implementations as well; a large number of MUXs in virtually all cyclic-locked benchmarks that have been provided by the original authors online exhibit this fault~\cite{TrustHubLink}. The same is true for the cyclic locking that is available in the same author's tool NEOS~\cite{NEOS}. This fault enables the application of SAAM. This problem is somewhat resolvable if dummy gates are deployed; however, this requires (trusted) layout manipulations to be cost-effective~\cite{cyclicObf2017}. The second structural fault hides in the insertion of \emph{two} MUXs when the fan-out of a gate equals 1. Since $w_{2}$ is the only wire that simultaneously drives \emph{both} MUXs, according to the scheme, it must be the \emph{true} wire of the MUX placed in the cycle. Hence, both MUXs can be removed. This issue persists even if both MUXs have separate key bits. Third, since a specific cycle length is induced by the scheme, in case a cycle is identified in the netlist, there exists a high likelihood that the longest backward edge is the \emph{false} one. Here, the length can be measured in terms of how many gates the edge "jumps over" from its \emph{first} entry point to the \emph{last} exit point in a \emph{topological} order. For example, topologically (when ignoring backward edges) the first entry point in Fig.~\ref{fig:cyclic_obf} is $M_{0}$ and the last exit point is $G_{3}$. Hence, the backward edge from $w_{3}$ to $w_{0}$ is clearly the longest one and its length is \emph{defined} as an input parameter. Finally, another issue occurs when not enough false wires are available (verified with NEOS). Here, the algorithm inserts a MUX that is driven by two inputs that originate from the \emph{same} MUX. Evidently, this structural composition immediately reveals the correct key-bit value. Consequently, due to the presented faults, cyclic obfuscation fails both ANT and RNT. To the best of our knowledge, the introduced faults have not been discussed before. Hence, we urge the authors to address these issues in the implementation and the theoretical work. Some faults can be patched by utilizing the D-MUX locking strategies. Moreover, based on the presented concepts of cyclic locking, multiple next-generation cyclic LL schemes have been proposed~\cite{SRCLock2018,8341984,looplock2021}, some of which try to mitigate the shortcomings of the initial proposal. However, using the presented evaluation concepts, a detailed analysis might reveal new structural faults in these schemes as well. Hence, we leave this task for future work.









\begin{figure}[t]
	\centering
	\subfloat[Original netlist]{
		\includegraphics[valign=c,width=0.3\columnwidth]{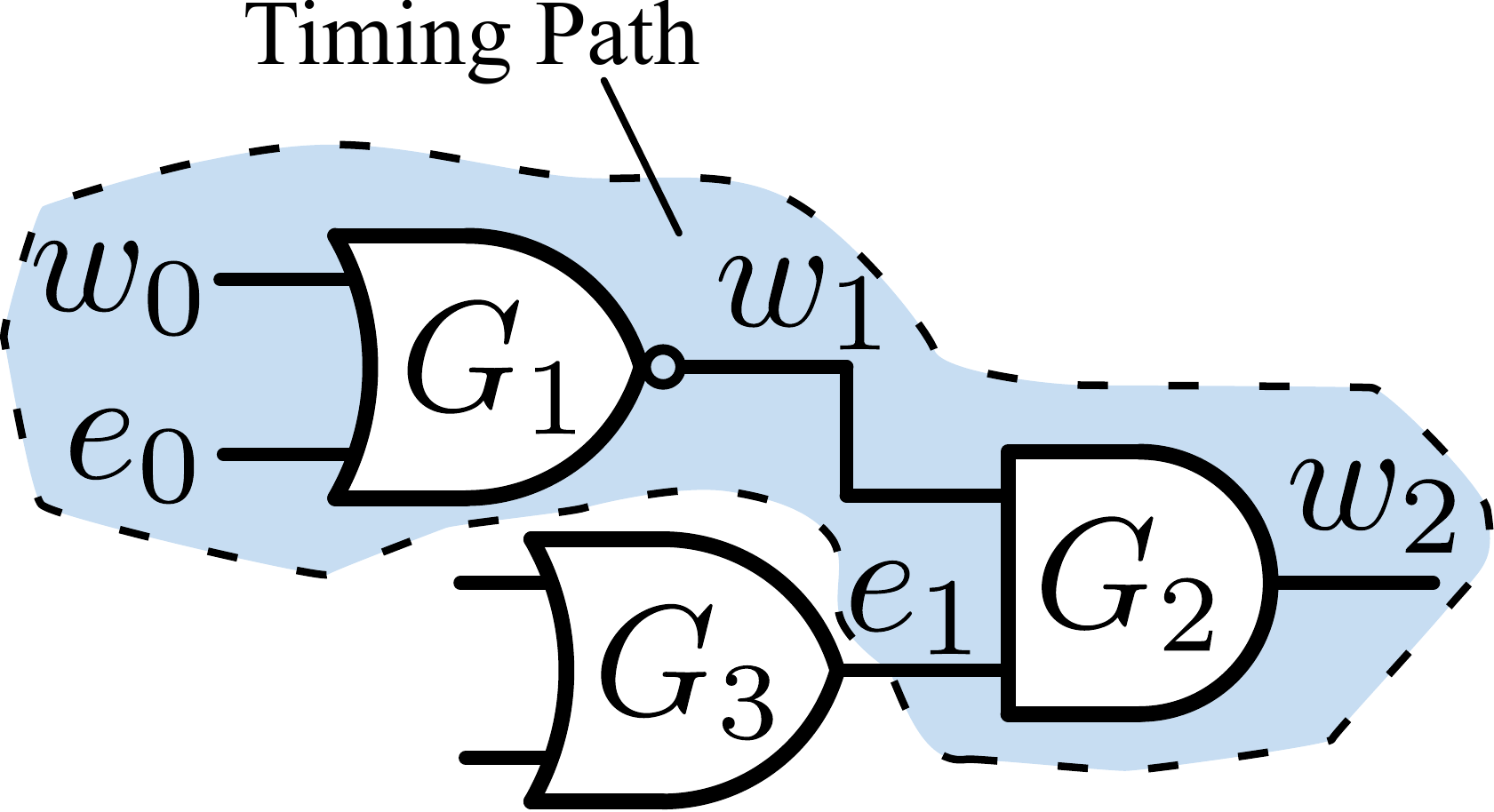}
		\vphantom{\includegraphics[width=0.095\columnwidth,valign=t]{example-image-10x16}}%
	}
	\subfloat[KeyRB]{
		\includegraphics[valign=c,width=0.54\columnwidth]{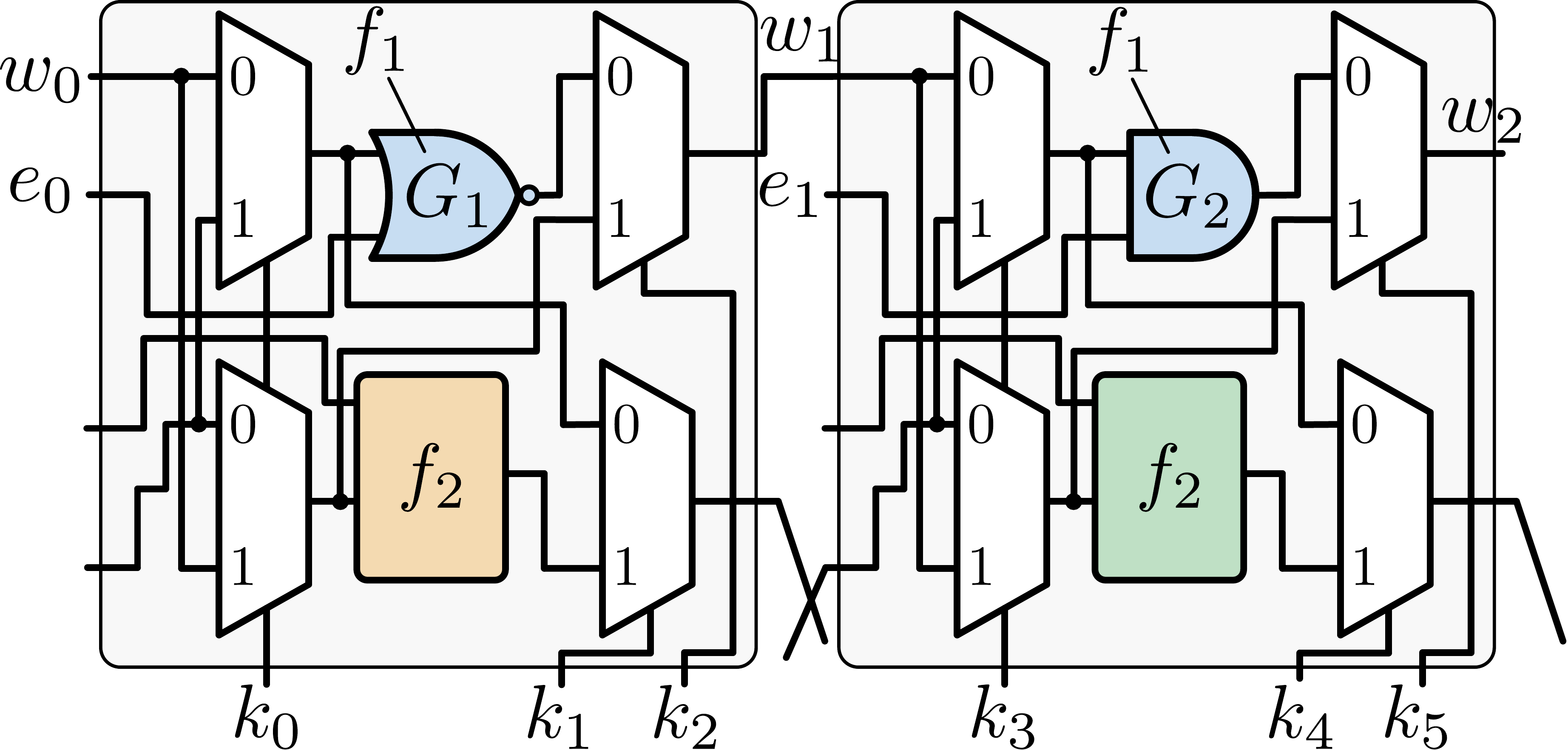}
	}
	\caption{InterLock: embedding a timing path into a KeyRB.}
	\label{fig:interlock}
\end{figure}

\textbf{Routing-Based LL:}~A promising approach to thwarting structural attacks lies in a new class of LL policies known as routing-based obfuscation. Even though these techniques have been designed to tackle SAT-based attacks, their structural composition might offer a fruitful ground to protect against structural ML-based attacks. The main building blocks of routing-based LL are \emph{key-programmable routing blocks} (keyRBs), which can implement a variety of topologies. For example, InterLock~\cite{InterLock2020} utilizes keyRBs that are composed of Switch-Boxes (SwBs) to embed selected timing paths of a predefined length. For example, the marked path in Fig.~\ref{fig:interlock}~(a) can be embedded into the keyRB by connecting multiple SwBs as shown in Fig.~\ref{fig:interlock}~(b). Each SwB hosts two 2-input gates ($f_{1}$ and $f_{2}$). These represent original gates from the netlist. For a correct key ($\{k_{0},k_{2},k_{3},k_{4}\}=\{0,0,0,0\}$), the embedded gates remain properly connected. 
Similar to D-MUX, this mechanism can create paths that are equally likely for each key value. However, some structural cases remain unresolved. For example, for $\{k_{1},k_{2}\}=\{0,1\}$, both functions become dangling gates, i.e., $e_{0}$ might lead to an \emph{unconnected} gate; thus suggesting that these key combinations are not valid. This problem is further exacerbated in case any wire in the selected path drives multiple gates. For example, if $w_{1}$ is connected to another gate (besides $G_{2}$), it is clear that $\{k_{1},k_{2}\}=\{0,1\}$ may not appear. These cases have not been clarified in the proposed work~\cite{InterLock2020}. Therefore, there is still the possibility to resolve such structural cases within InterLock. Moreover, we see a good potential to reuse the locking strategies presented in D-MUX for furthering the concept of InterLock. Thus, we leave this to future work.

A similar scheme is known as Banyan locking~\cite{BanyanLocking2020}. This scheme expands the SwBs by adding decoy logic. For example, the SwB can be implemented as a 2-input-2-output function, where each output can be driven by the original gate, decoy gate, buffered input wire, or a constant. This configuration leads to potential structural concerns. For example, in ANT, all SwBs would contain only AND gates, thus reducing the key space as multiple keys would result in the same output. Thus, Banyan LL might result in structural leakage depending on the structural characteristics of the netlist. Moreover, as Banyan LL implements a set of gates in addition to multiple N-to-1 MUXs for a single SwB, its overhead is likely to be a major concern for wide adoption. Even its authors left the cost evaluation to future work~\cite{BanyanLocking2020}. Nevertheless, similar to InterLock, there is a potential to adapt Banyan LL based on the introduced D-MUX locking strategies.

\textbf{LUT-Based LL:}~A promising concept of LL is manifest in LUT-based locking~\cite{logicBariers2010,LUTLock2018}. This LL type replaces parts of the target design with generic lookup tables that are configured \emph{after} fabrication. This approach has one major advantage compared to traditional LL policies; the LUTs exhibit a generic memory layout topology, regardless of their functionality. Thus, if LUT-based LL is deployed, part of the design is not available to the attacker (e.g., the untrusted foundry); resulting in a structurally secure design. However, this feature comes with a few disadvantages. First, utilizing LUTs for LL can lead to considerable overheads, especially in the traditional CMOS technology. Second, LUT-based LL requires the inclusion of a trusted testing facility, which might be a concern for some IP owners. Third, to lower the cost aspects, the integration of novel technologies is required (such as the non-volatile spin transfer torque magnetic technology). Hence, another burden is put on the design flow. Nevertheless, as LUTs are able to absorb parts of the design \emph{without leaving LL-induced structural traces}, they provide a promising direction to design resilient schemes. Interestingly, ANT provides an important observation; if the structure of the netlist is very regular (e.g., AND tree), the LUTs must absorb \emph{more than one gate} to not leak information; otherwise it is clear what functionality is implemented. This can be simulated by generalizing the Twin-Gate scheme to an N-Gate scheme, where N is the number of all possible two-input gates. This suggests that the circuit family might play a role in LUT-based LL. 

\textbf{ML-Resilient LL:}~UNSAIL inserts key gates with the goal to create confusing training data for ML-based attacks (such as SAIL)~\cite{unsail2020}. For a selected LL policy, UNSAIL locks the netlist using half of the available key. Next, the design is synthesized. At this point, a dictionary of observations is generated by extracting the key-gate subgraphs from both locked designs \emph{before} and \emph{after} synthesis. Next, UNSAIL performs a dictionary-guided key-gates insertion, thereby using the remaining half of the key. Thus, UNSAIL-locked netlists ensure that equivalent subgraphs are linked to different key values; resulting in confusing training data. Even though deceptive observations are crucial to thwarting ML-based attacks, we can identify a few disadvantages of UNSAIL. First, the major issue of UNSAIL is its deep reliance on the usage of a synthesis tool. As discussed in Sections~\ref{resynthesis-and-security} and~\ref{lessons-learned}, in the context of learning resilience, the security of LL should not depend on the specifics of synthesis transformations, especially since these are typically induced by closed-source, third-party software. However, note that often the synthesis tool \emph{is not} part of the attack model, i.e., it is regarded as trustworthy. Hence, the dependency on the tool must not be seen as a disadvantage in these cases. Second, in the optimal case, an UNSAIL-locked netlist should not be changed through additional synthesis. Nevertheless, resynthesis and layout generation might lead to new leakage points that reverse the effects of UNSAIL. Moreover, tampering with the synthesis can result in suboptimal designs. Third, the resiliency of UNSAIL has only been demonstrated for schemes that buffer or invert a single wire, such as XOR/XNOR-locking and single-wire MUX gates (MUXs that are driven by the same wire twice, except that one input is inverted). These schemes fail the theoretical tests. Finally, UNSAIL only lowers the accuracy of the attacks for a certain amount of percentage points. In comparison, D-MUX is independent of synthesis transformations and exhibits maximal resilience against the introduced attacks.

A potentially ML-resilient scheme is known as Scalable Attack-Resistant Obfuscation (SARO)~\cite{alaql2020scalable}; operating in two steps. First, it splits the design into smaller partitions with the aim to maximize structural netlist changes. Second, SARO deploys a Truth Table Transformation (T3) to lock the created partitions. T3 is designed to induce randomized design alterations, thus increasing the complexity of deploying pattern-recognition attacks (such as SAIL). Even though the effectiveness of SARO has been evaluated against a variety of attacks, its resilience against ML-based attacks is only assumed based on a proposed metric that tries to capture the level of functional and structural changes in the design. Thus, its effectiveness against ML-based attacks remains to be evaluated.




\section{Limitation and Opportunities}\label{limitations}
Even though we could not identify any leakage in D-MUX, this does not conclusively prove its absence. Thus, D-MUX has to be further evaluated against upcoming attacks~\cite{gnnunlock2020}.


Another interesting problem lies in the evaluation of structural leakage that is beyond the capability of the human observer. For example, LL might leak key-related information only on a \emph{topological} level. Evidently, comparing the topology of netlists is not easily done for humans. Moreover, this problem has not been addressed yet in detail as LL mostly induces local changes. Nevertheless, the proposed tests offer the capability to pinpoint fundamental security vulnerabilities that are easily exploitable. Whether a global leakage exists will require the inclusion of suitable topology-matching procedures, hence opening up challenging questions in LL.

We have seen that the security of LL is often predicated by the structural characteristics of the target netlists, i.e., the circuit family. However, a comprehensive analysis of the existence of a structural bias across a wide range of designs has yet to be performed. A first indication is provided by the success of SnapShot~\cite{sisejkovic2020challenging}. Here, a range of benchmarks was successfully attacked, suggesting that there exists a bias towards repeating structures in hardware designs. The results of such a study could provide some insights on what LL should focus on---in case a general solution is difficult to provide.

Finally, we have identified the potential to transfer the D-MUX strategies to OG-resilient LL, such as routing-based LL.
%
%
%

\section{Conclusion}\label{conclusion}
This work addresses the challenges of logic locking in the context of learning-based attacks. Hereby, the first theoretical concept for evaluating learning resilience in logic locking has been introduced; exposing critical information leaks even before concrete locking implementations are in place. Based on the theoretical insights, we developed D-MUX; the first deceptive logic-locking scheme, thereby investigating and correcting a major fallacy in existing MUX-based locking. 
The resiliency of D-MUX was evaluated through the extensive application of the latest machine-learning-based and MUX-targeting attacks. The cost of D-MUX has been modeled theoretically and evaluated using a concrete technology node.
Finally, we have discussed related locking concepts, thereby providing novel insights into potential information leakage as well as promising research directions. With the presented work, we establish the cornerstones for the design of next-generation logic locking in the era of machine learning.

\ifCLASSOPTIONcaptionsoff
\newpage
\fi

\bibliographystyle{IEEEtran}
\bibliography{bibliography_full_short}

\vspace{-0.4in}
\begin{IEEEbiography}[{\includegraphics[width=1in,height=1.25in,clip,keepaspectratio]{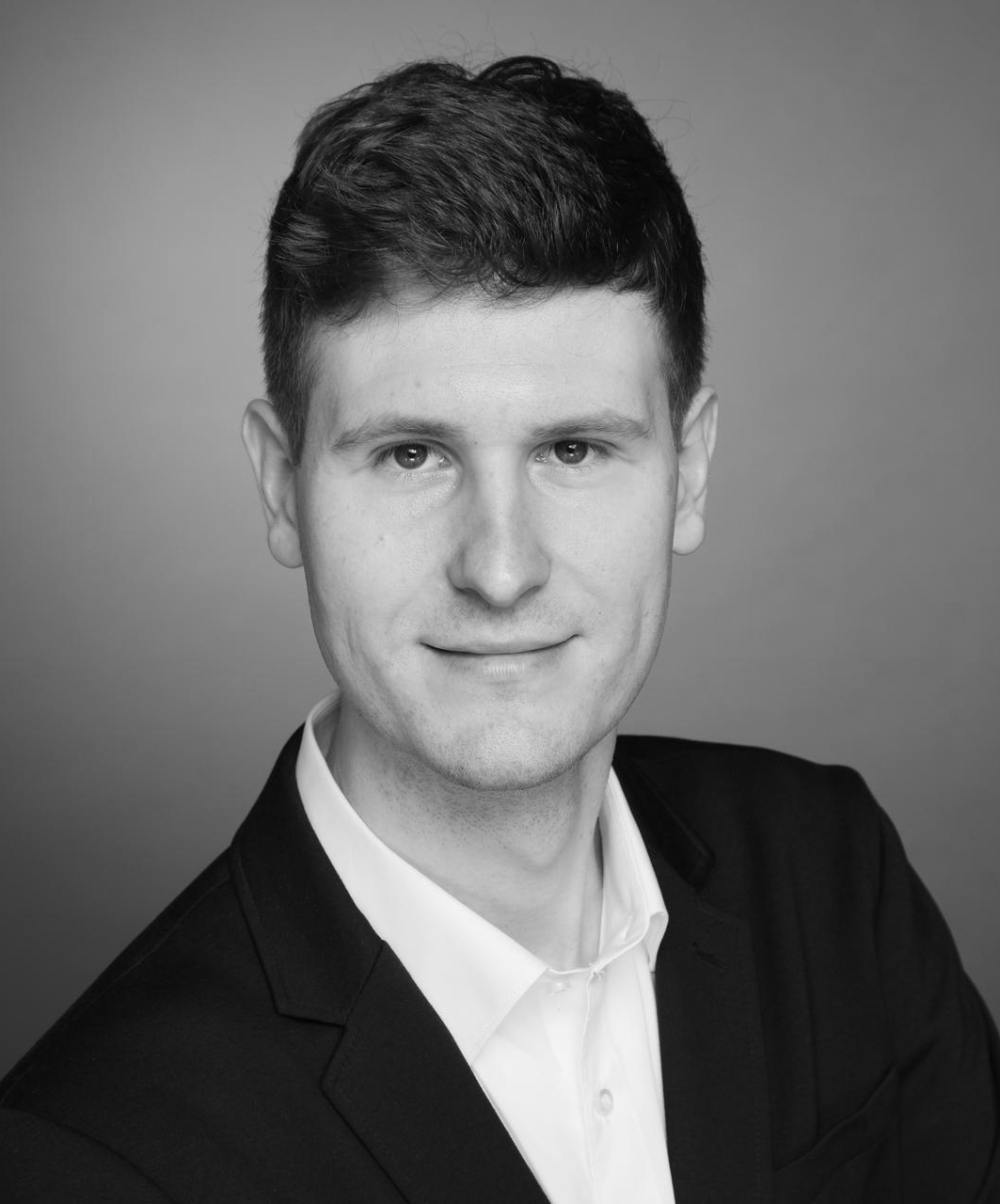}}]{Dominik Sisejkovic} received the B.Sc. and M.Sc. degree in computing from the University of Zagreb, Croatia, in 2014 and 2016 respectively. In 2016, he started working as a Ph.D. student and research assistant at RWTH Aachen University. His research interest includes hardware security and machine learning for security. He was directly involved in the design and implementation of the logic-locking framework that was applied for the production of the first logic locked RISC-V processor on the market.
\end{IEEEbiography}
\vspace{-0.5in}
\begin{IEEEbiography}[{\includegraphics[width=1in,height=1.25in,clip,keepaspectratio]{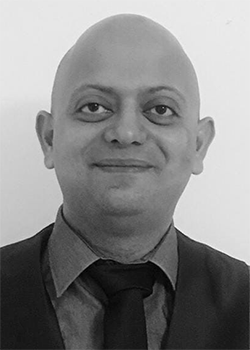}}]{Farhad Merchant}
received his Ph.D. from the Indian Institute of Science, Bangalore (India), in 2016. His Ph.D. thesis title was "Algorithm-Architecture Co-design for Dense Linear Algebra Computations". He worked as a postdoctoral research fellow at NTU Singapore, from March 2016 to December 2016.  He joined Institute for Communication Technologies and Embedded Systems, RWTH Aachen University, in December 2017 as a postdoctoral research fellow in the Chair for Software for Systems on Silicon.
\end{IEEEbiography}
\vspace{-0.5in}
\begin{IEEEbiography}[{\includegraphics[width=1in,height=1.25in,clip,keepaspectratio]{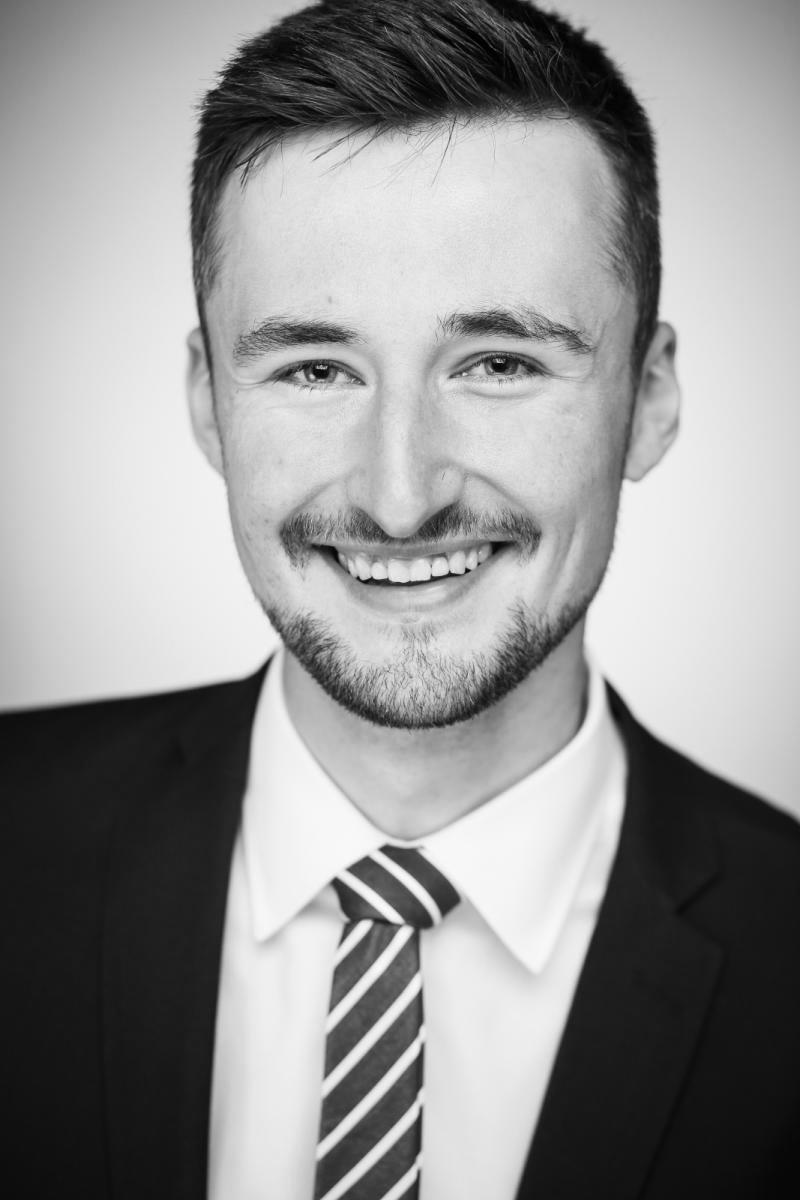}}]{Lennart Reimann}
received his Bachelor's and Master's degree in Electrical Engineering, Information Technology, and Computer Engineering from RWTH Aachen University in 2016 and 2019, respectively. After his Master's, he started working toward his Ph.D. as a research assistant at ICE under the supervision of Prof. Leupers. His research interest includes Hardware Security, Secure ASIP design, Cryptographic accelerator design, etc.
\end{IEEEbiography}
\vspace{-0.4in}
\begin{IEEEbiography}[{\includegraphics[width=1in,height=1.25in,clip,keepaspectratio]{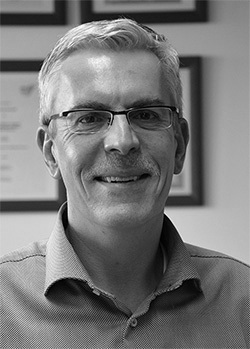}}]{Rainer Leupers}
received the M.Sc. (Dipl.-Inform.) and Ph.D. (Dr. rer. nat.) degrees in Computer Science with honors from TU Dortmund in 1992 and 1997. From 1997-2001 he was the Chief Engineer at the Embedded Systems Chair at TU Dortmund. In 2002, he joined RWTH Aachen University as a professor for Software for Systems on Silicon. His research comprises embedded software development tools, multicore processor architectures, hardware security, and system-level electronic design automation. 
\end{IEEEbiography}
\vfill





\end{document}